\documentclass[%
 reprint,%
 aip,cha,%
]{revtex4-1}

\usepackage{docs}%
\usepackage[colorlinks=true,linkcolor=blue]{hyperref}%

\usepackage{epsfig}
\usepackage{url}
\usepackage{amssymb, amsmath, amsthm}
\usepackage{epsfig,graphicx}
\usepackage{xspace}
\usepackage{color}
\usepackage{verbatim}
\usepackage{pifont} % check and x marks
\usepackage{color}
\usepackage{listings}
\usepackage{url}
\usepackage{amssymb, amsmath, amsthm}
\usepackage{epsfig,graphicx}
\usepackage{tikz}
\usetikzlibrary{shapes,backgrounds}
\usepackage{subcaption}
\usepackage{xspace}
\usepackage{color}
\usepackage{verbatim}
\usepackage{pifont} % check and x marks
\usepackage{color}
\usepackage{listings}
\usepackage{mathtools} % mathclap

\begin{document}

\title{
%Using stored predictive information to optimize delay-coordinate reconstruction parameter selection for understanding the future \\
% Optimal delay reconstructions for time-series forecasting  (too broad)
A new method for choosing parameters in delay reconstruction-based forecast strategies
% Independent coordinates for delay reconstruction-based forecast strategies (a la F&S)
}

\author{Joshua Garland}%
\email{joshua.garland@colorado.edu}
\affiliation{University of Colorado\\Department of Computer Science\\ Boulder, Colorado 80303, USA}%
\author{Ryan G. James}%
\email{rgjames@ucdavis.edu}
\affiliation{University of California\\Department of Physics, Davis, California 95616, USA}%
\author{Elizabeth Bradley}%
\email{lizb@colorado.edu}
\affiliation{University of Colorado\\Department of Computer Science, Boulder, Colorado 80303, USA \\ and the Santa Fe Institute, Santa Fe, New Mexico}%
%\date{}%
%\revised{}%

\maketitle

%Editing Commands
\newcommand{\cmark}{\ding{51}}
\newcommand{\xmark}{\ding{55}}
\newcommand{\alert}[1]{{\color{red}#1}}
\newcommand{\ftnote}[1]{{\color{orange}#1}}
\newcommand{\espace}[2]{\ensuremath{\mathbb{E}[#1,#2]}}
\newcommand{\mytau}{SPI\xspace}

\newcommand{\cmt}[1]{\textcolor{magenta}{#1}}

%Theorem Commands
\newtheorem*{mydef}{Definition}
\newtheorem{thm}{Theorem}
\newtheorem*{remark}{Remark}
\newtheorem{prop}[thm]{Proposition}
\newtheorem{example}[thm]{Example}

%Program commands

\newcommand{\gcc}{{\tt 403.gcc}\xspace}
\newcommand{\sphinx}{{\tt 482.sphinx}\xspace}
\newcommand{\eig}{{\tt dgeev}\xspace}

\newcommand{\naive}{na\"ive\xspace}
\newcommand{\col}{{\tt col\_major}\xspace}

%I-Diagram Setup
\usetikzlibrary{calc}
\tikzset{filled/.style={fill=blue, opacity=0.2}}

\def \setupdiagrams {
  \def \delt {2cm}
  \def \radius {2cm}

  \coordinate (A) at (0, 0);
  \coordinate (B) at (240:\delt);
  \coordinate (C) at (-60:\delt);

  \coordinate (center) at (barycentric cs:A=1/3,B=1/3,C=1/3);

  \def \Acirc { (A) circle (\radius) }
  \def \Bcirc { (B) circle (\radius) }
  \def \Ccirc { (C) circle (\radius) }

  \def \drawdiagram {
    \draw[very thick] \Acirc;
    \draw[very thick] \Bcirc;
    \draw[very thick] \Ccirc;

    \node at ($ (center) ! 3   ! (A) $) {$H[X_{j+p}]$};
    \node at ($ (center) ! 3.5 ! (B) $) {$H[X_{j-\tau}]$};
    \node at ($ (center) ! 3.5 ! (C) $) {$H[X_{j}]$};
  }
}

%citation commands
\newcommand{\citeNTau}{\cite{fraser-swinney,Olbrich97,kantz97,Buzug92Comp,liebert-wavering,Buzugfilldeform,Liebert89,rosenstein94}\xspace}

\newcommand{\citeNM}{\cite{liebert-wavering,Cao97Embed,Kugi96,Buzugfilldeform,KBA92,Hegger:1999yq,kantz97}\xspace}

\newcommand{\citeNEPARAMS}{\cite{Olbrich97,fraser-swinney,kantz97,Buzug92Comp,liebert-wavering,Buzugfilldeform,Liebert89,rosenstein94,Cao97Embed,Kugi96,KBA92,Hegger:1999yq}}

\newcommand{\citeDCEFORECASTING}{\cite{weigend-book,casdagli-eubank92,Smith199250,pikovsky86-sov,sugihara90,lorenz-analogues}}

%Forecast Commands
%\newcommand{\kfnnLMA}{{\tt $k$-ball fnn-LMA}\xspace}
%\newcommand{\kroLMA}{{\tt $k$-ball ro-LMA}\xspace}

\newcommand{\fnnLMA}{{\tt fnn-LMA}\xspace}
\newcommand{\roLMA}{{\tt ro-LMA}\xspace}
\newcommand{\ipc}{{\tt IPC}\xspace}

\section*{Abstract}

Delay-coordinate reconstruction is a proven modeling strategy for
building effective forecasts of nonlinear time series. The first step
in this process is the estimation of good values for two parameters,
the time delay and the embedding dimension.  Many heuristics and
strategies have been proposed in the literature for estimating these
values.  Few, if any, of these methods were developed with forecasting
in mind, however, and their results are not optimal for that purpose.
Even so, these heuristics---intended for other applications---are
routinely used when building delay coordinate reconstruction-based
forecast models.  In this paper, we propose a new strategy for
choosing optimal parameter values for forecast methods that are based
on delay-coordinate reconstructions.  The basic calculation involves
maximizing the shared information between each delay vector and the
future state of the system.  We illustrate the effectiveness of this
method on several synthetic and experimental systems, showing that
this metric can be calculated quickly and reliably from a relatively
short time series, and that it provides a direct indication of how
well a near-neighbor based forecasting method will work on a given
delay reconstruction of that time series.  This allows a
practitioner to choose reconstruction parameters that avoid any
pathologies, regardless of the underlying mechanism, and maximize the
predictive information contained in the reconstruction.

%\chapter{ }\label{ch:methods}

\section{Introduction}
\label{sec:mytaumotivation}

The method of delays is a well-established technique for
reconstructing the state-space dynamics of a system from scalar
time-series data\cite{takens,packard80,sauer91}.  The task of choosing
good values for the free parameters in this procedure has been the
subject of a large and active body of literature over the past few
decades, e.g.,\citeNEPARAMS.  The majority of these techniques focus
on the geometry of the reconstruction.  A standard method for
selecting the delay $\tau$, for instance, is to maximize independence
between the coordinates of the delay vector while minimizing
overfolding and reduction in causality between
coordinates\cite{fraser-swinney}; a common way to choose an embedding
dimension is to track changes in near-neighbor relationships in
reconstructions of different dimensions\cite{KBA92}.

This heavy focus on the geometry of the delay reconstruction is
appropriate when one is interested in quantities like fractal
dimension and Lyapunov exponents, but it is not necessarily the best
approach when one is building a delay reconstruction \emph{for the
  purposes of prediction}.  That issue, which is the focus of this
paper, has received comparatively little attention in the extensive
literature on delay reconstruction-based
prediction\citeDCEFORECASTING.  In the following section, we propose a
robust, computationally efficient method called \mytau that can be
used to select parameter values that maximize the shared information
between the past and the future---or, equivalently, that maximize the
reduction in uncertainty about the future given the current model of
the past.  The implementation details, and a complexity analysis of
the algorithm, are covered in Section~\ref{sec:implementation}.  In
Section~\ref{sec:results}, we show that simple prediction methods
working with \mytau-optimal reconstructions---constructions using
parameter values that follow from the \mytau calculations---perform
better, on both real and synthetic examples, than those same forecast
methods working with reconstructions that are built using the
traditional methods mentioned above. Finally, in
Section~\ref{sec:dataandhorizon} we explore the utility of \mytau in
the face of different data lengths and prediction horizons.
% In Section~\ref{sec:2D-example}, we extend that exploration to
%reconstructions that use fewer dimensions than the
%theoretical\cite{embedology} or practical\cite{KBA92} requirements on
%that parameter.
%
% \alert{we are out of time to do this in the arxiv version, maybe in the real paper and maybe Ryan can help do this...: For both types of reconstructions, we show that computations of these \mytau-optimal parameter values are quite robust to noise and short data lengths, and that the results of these calculations are identical to the true optimum values, as determined by exhaustive search.  Exhaustive search is of course not practical in a real prediction application, so having a way to estimate good parameter values quickly and robustly is a significant advantage.}

\section{Shared information and delay reconstructions}
\label{sec:sharedinfo}

The information shared between the past and the future is known as the
excess entropy\cite{crutchfield2003regularities}.  We will denote it
here by $E = I[\overleftarrow{X};\overrightarrow{X}]$, where $I$ is
the mutual information\cite{yeung2012first} and $\overleftarrow{X}$
and $\overrightarrow{X}$ represent the infinite past and the infinite
future, respectively.  $E$ is often difficult to estimate from data
due to the need to calculate statistics over potentially infinite
random variables\cite{james2014many}.  While this is possible in
principle, it is too difficult in practice for all but the simplest of
dynamics\cite{strelioff2014bayesian}.  In any case, the excess entropy
is not exactly what one needs for the purposes of prediction, since it
is not realistic to expect to have the infinite past or to predict
infinitely far into the future.  For our purposes, it is more
productive to consider the information contained in the \emph{recent}
past and determine how much that explains about the not-too-distant
future.  To that end, we define
\begin{align*}
  \mytau={I}[\mathcal{S}_j;X_{j+p}]
  ~,
\end{align*}
where $\mathcal{S}_j$ is an estimate of the state of the system at
time $j$ and $X_{j+p}$ is the state of the system $p$ steps in the
future.

\begin{figure}
  \centering
  \begin{tikzpicture}[baseline=0]
    \setupdiagrams

    \begin{scope}
      \clip \Bcirc;
      \fill[filled] \Ccirc;
    \end{scope}

    \drawdiagram

  \end{tikzpicture}
  \caption{An I-diagram of the time-delayed mutual information.  The
    circles represent uncertainties ($H$) in different variables; the
    shaded region represents $I[X_{j};X_{j-\tau}]$, the time-delayed
    mutual information between the current state $X_{j}$ and the state
    $\tau$ time units in the past, $X_{j-\tau}$.  Notice that the shaded
    region is indifferent to $H[X_{j+p}]$, the uncertainty about the future. 
%    it will always be however much $H[X_{j-\tau}]$ and $H[X_{j}]$
%    overlap, regardless of how much $H[X_{j+p}]$ overlaps, or doesn't,
%    with the other two.
}
  \label{fig:MI-I-Diagram}
\end{figure}
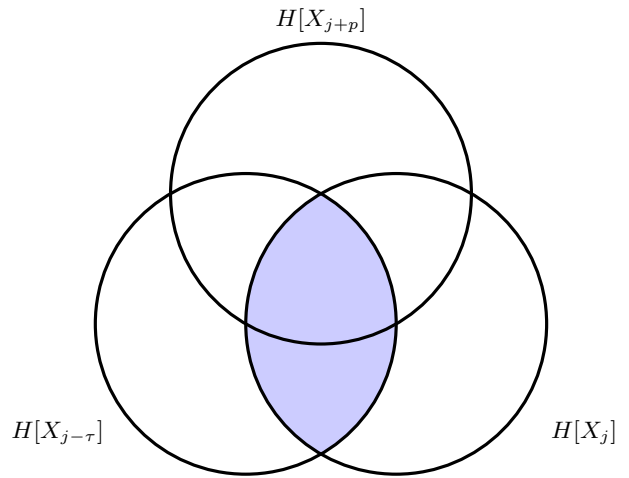

This can be neatly visualized---and compared to traditional methods
like time-delayed mutual information, multi-information and the
so-called co-information\cite{Bell03theco-information}---using the
I-diagrams of
Yeung\cite{yeung2012first}. Figure~\ref{fig:MI-I-Diagram} shows an
I-diagram of time-delayed mutual information for a specific $\tau$.
In a diagram like this, each circle represents the uncertainty in a
particular variable.  The left circle in
Figure~\ref{fig:MI-I-Diagram}, for instance, represents the average
uncertainty in observing $X_{j-\tau}$ (i.e., $H[X_{j-\tau}]$, where
$H$ is the Shannon entropy\cite{yeung2012first}); similarly, the top
circle represents $H[X_{j+p}]$ or the uncertainty in the $p^{th}$ future
observation.  Each of the overlapping regions represents shared
uncertainty: e.g., in Figure~\ref{fig:MI-I-Diagram}, the shaded region
represents the shared uncertainty between $X_{j}$ and $X_{j-\tau}$.
More precisely, the shaded region schematizes the quantity
\begin{eqnarray*}
  I[X_{j};X_{j-\tau}] &=& H[X_{j}] + H[X_{j-\tau}] - H[X_{j},X_{j-\tau}] \\
  &=&H[X_{j}] - H[X_{j}|X_{j-\tau}] \\
  &=& H[X_{j-\tau}] - H[X_{j-\tau}|X_{j}].
\end{eqnarray*}

If the $X$ are trajectories in reconstructed state space, then tuning
the reconstruction parameters (e.g., $\tau$) changes the size of the
overlap regions---i.e., the amount of information shared between the
coordinates of the delay vector.  This notion can be put into practice
to select good values for those parameters.  Notice, for instance,
that minimizing the shaded region in
Figure~\ref{fig:MI-I-Diagram}---that is, rendering $X_{j}$ and
$X_{j-\tau}$ as independent as possible---maximizes the total
uncertainty that is explained by the combined model
$[X_{j},X_{j-\tau}]^T$ (the sum of the area of the two circles).  This
is precisely the argument made by Fraser and Swinney in
\cite{fraser-swinney}.  However, it is easy to see from the I-diagram
that choosing $\tau$ in this way does not explicitly take into account
explanations of the \emph{future}---that is, it does not reduce the
uncertainty about $X_{t+p}$. Moreover, the calculation does not extend
to three or more variables, where minimizing overlap is not a trivial
extension of the reasoning captured in the I-diagrams.

The obvious next step would be to explicitly include the future in the
estimation procedure.  One approach to this would be to work with the
so-called co-information\cite{Bell03theco-information},
\begin{align*}
  \mathcal{C}= I[X_{j};X_{j-\tau};X_{j+p}]
  ~,
\end{align*}
As depicted in Figure~\ref{fig:co-information}, this is the intersection of
$H[X_{j}]$, $H[X_{j-\tau}]$ and $H[X_{j+p}]$.  It describes the
reduction in uncertainty that the \emph{two} past states, together,
provide regarding the future.  While this is obviously an improvement
over the time-delayed mutual information of
Figure~\ref{fig:MI-I-Diagram}, it does not take into account the
information that is shared between $X_j$ and the future but \emph{not
  shared with the past} (i.e., $X_{j-\tau}$), and vice versa.  The
so-called multi-information,
\begin{align*}
  \mathcal{M} = \sum_{\mathclap{i \in \{j, j-\tau, j+p\}}} \left( H[X_i] \right) - H[X_{j},X_{j-\tau},X_{j+p}]
  ~,
\end{align*}
depicted in Figure~\ref{fig:multi-information} addresses this
shortcoming, but it also includes information that is shared between
the past and the present, but not with the future.  This is not
terribly useful for the purposes of prediction.  Moreover, the
multi-information overweights information that is shared between all
three circles---past, present, and future---thereby artificially
over-valuing information that is shared in all delay coordinates.  In
the context of predicting $X_{t+p}$, the provenance of the information
is irrelevant and so the multi-information seems ill-suited to the
task at hand as well.

\begin{figure}[ht!]
  \begin{subfigure}[b]{0.49\textwidth}
    \begin{tikzpicture}[baseline=0,label=$m$]
      \setupdiagrams

      \begin{scope}
        \clip \Acirc;
        \clip \Bcirc;
        \fill[filled] \Ccirc;
      \end{scope}

      \drawdiagram
    \end{tikzpicture}
    \caption{The co-information, $\mathcal{C}[X_{j+1};X_j;X_{j-\tau}]$}
    \label{fig:co-information}
  \end{subfigure}%
  \qquad
  \begin{subfigure}[b]{0.49\textwidth}
    \begin{tikzpicture}[baseline=0]
      \setupdiagrams

      \begin{scope}[even odd rule]
	      \clip \Acirc;
        \clip \Acirc \Bcirc;
        \fill[filled] \Ccirc;
      \end{scope}

      \begin{scope}[even odd rule]
	      \clip \Ccirc;
        \clip \Ccirc \Acirc;
        \fill[filled] \Bcirc;
      \end{scope}

      \begin{scope}[even odd rule]
	      \clip \Bcirc;
        \clip \Bcirc \Ccirc;
        \fill[filled] \Acirc;
      \end{scope}

      \begin{scope}
        \clip \Acirc;
        \clip \Bcirc;
        \fill[filled] \Ccirc;
        \fill[filled] \Ccirc;
      \end{scope}

      \drawdiagram
    \end{tikzpicture}
    \caption{An I-diagram of the multi-information,
      $\mathcal{M}[X_{j},X_{j-\tau};X_{j+p}]$. The centermost
      region is more darkly shaded here to reflect the extra weight
      that that region carries in the calculation.}
    \label{fig:multi-information}
  \end{subfigure}
%  \caption{Two possible generalizations of the mutual information.}
\end{figure}

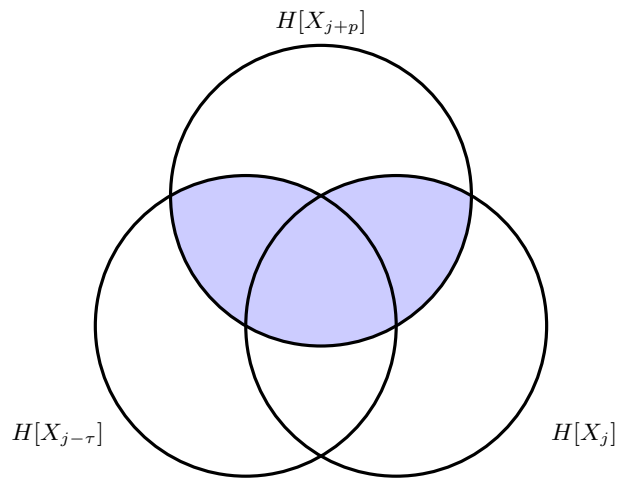
\begin{figure}[ht!]
\centering
    \begin{tikzpicture}[baseline=0]
      \setupdiagrams

      \begin{scope}
        \clip \Bcirc \Ccirc;
        \fill[filled] \Acirc;
      \end{scope}

      \drawdiagram
    \end{tikzpicture}
  \caption{An I-diagram of \mytau, the quantity proposed in this
    paper: $I[[X_{j},X_{j-\tau}];X_{j+p}]$.  This quantity captures
    the shared information between the past, present, and future
    independently, as well as the information that the past and
    present, together, share with the future.}
  \label{fig:mytau}
\end{figure}

\mytau addresses all of the issues raised in the previous
paragraphs. By treating the generic delay vector as a joint variable,
rather than a series of single variables, \mytau captures the shared
information between the past, present, and future independently (the
left and right colored wedges in Figure~\ref{fig:mytau}), as well as
the information that the past and present, together, share with the
future (the center wedge).  By choosing delay reconstruction
parameters that maximize \mytau, then, one can explicitly maximize the
amount of information that each delay vector contains about the
future.

To make all of this more concrete and tie it back to state-space
prediction of dynamical systems, consider the following example: let
$\mathcal{S}_j$ be a two-dimensional delay reconstruction of the time
series, $\mathcal{S}_j = [x_{j},x_{j-\tau}]^T$.  In this case, \mytau
becomes $I[[X_{j},X_{j-\tau}]^T;X_{t+p}]$, which describes the
reduction in uncertainty about the system at time $j+p$, given the
state estimate $[X_{j},X_{j-\tau}]^T$.  One can estimate a $\tau$ value for
the purposes of reconstructing the dynamics from a given time series,
for instance, by calculating \mytau for a range of $\tau$ and choosing
the first maximum (i.e., minimizing the uncertainty about the $p^{th}$
future observation).  One can then apply any state-space forecasting
method to the resulting reconstruction in order to predict the future
course of that time series. In Section~\ref{sec:results}, we explore
that claim using Lorenz's classic method of
analogues\cite{lorenz-analogues}, but it should be just as applicable for
other predictors that utilize state-space reconstructions, such as the methods used in\cite{weigend-book,casdagli-eubank92,Smith199250,sugihara90}.
%\alert{That claim is both broad and imprecise.  Which methods are in
%  that ``most'' and which ones aren't?  Why?}

Notice that both the definition of \mytau and its use in optimizing
forecast algorithms are general ideas that are easily extensible to
other state estimators. For example, in the case of traditional
delay-coordinate \emph{embedding}, the state estimator is the $m$-dimensional
delay vector, i.e.,
\begin{align*}
  \mathcal{S}_j = [X_{j},X_{j-\tau},\dots,X_{j-(m-1)\tau)}]^T
\end{align*}
with $m$ chosen to meet the appropriate theoretical requirements
\cite{takens,sauer91}. We demonstrate this approach in
Section~\ref{sec:results}.  If the time series is pre-processed (e.g., via a Kalman filter\cite{sorenson}, a low-pass
filter and an inverse Fourier transform\cite{sauer-delay}, or some
other local linear
transformation\cite{weigend-book,casdagli-eubank92,Smith199250,sugihara90,kantz97}
), the state estimator simply becomes $\mathcal{S}_j= \hat{\vec{x}}_j$
where $\hat{\vec{x}}_j$ is the processed $m$-dimensional delay vector.
As we demonstrate in Section~\ref{subsubsec:computer}, one can even
use \mytau to optimize parameter choices for forecast methods that use
reconstructions that are not embeddings---i.e., those whose dimensions
do not meet the traditional requirements for preserving dynamical
invariants like the Lyapunov exponent.

\section{Efficient estimation of \mytau}
\label{sec:implementation}

%\alert{
%\begin{enumerate}
%\item  estimating $\mathcal{I}$ from real-valued time series using the KSG %estimator
%\item JIDT and the algorithmic complexity of this.
%\item but that is ok because ...
%\item A nice illustration and discussion of how short the training signal (using 5\%% vs using the whole thing) for this can be showing this can be used on the fly even though it is sort of expensive compared to traditional binning and also explain the TISEAN implementation is purposefully very course and so gives a crazy estimate value wise but should be right shape but faster.
%\end{enumerate}
%}

To calculate \mytau from a real-valued time series, one must first symbolize those data.  Simple binning is not a good solution here, as it is known to cause severe bias if the bin boundaries do not create a generating partition\cite{KSG}.  A useful alternative is kernel estimation \cite{PhysRevLett.85.461,PhysRevLett.99.204101}, in which the relevant probability density functions are estimated via a function $\Theta$ with a resolution or bandwidth $r$ that measures the similarity between two points in $X \times Y$ space.  (For \mytau, $X$ would be $\mathcal{S}_j$ and $Y$ would be $X_{j+p}$.)  Given points $\{x_i,y_i\}$ and $\{x'_i,y'_i\}$ in $X \times Y$, one can define:
\begin{align*}
  \hat{p}_r(x_i,y_i) = \frac{1}{N} \sum_{i'=1}^N\Theta
  \left( \begin{array}{|c|}
    x_i-x'_i \\%
    y_i-y'_i
  \end{array} -r \right)
  ~,
\end{align*}
where $\Theta(x>0) = 0$ and $\Theta(x\le 0 ) = 1$.  That is,
$\hat{p}_r(x_i,y_i)$ is the proportion of the $N$ pairs of points in
$X\times Y$ space that fall within the kernel bandwidth $r$ of
$\{x_i,y_i\}$, i.e., the proportion of points similar to
$\{x_i,y_i\}$. When $|\cdot|$ is the max norm, this is the so-called box
kernel.  This too, however, can introduce bias\cite{jidt} and is
dependent on the choice of bandwidth $r$.  After these estimates, and the
analogous estimates for $\hat{p}(x)$, are produced, they are
then used directly to
compute local estimates of mutual information for each point in space,
which are then averaged over all samples to produce the mutual
information of the time series.  For more details on this procedure,
see\cite{jidt}.

A better way to calculate $I[X;Y]$ and estimate \mytau is the
Kraskov-St\"ugbauer-Grassberger (KSG) estimator\cite{KSG}.  This
approach dynamically alters the kernel bandwidth to match the density
of the data, thereby smoothing out errors in the probability density
function estimation process.  In this approach, one first finds the
$k^{th}$ nearest neighbor for each sample $\{x,y\}$ (using max norms
to compute distances in $x$ and $y$), then sets kernel widths $r_x$
and $r_y$ accordingly and performs the pdf estimation.  There are two
algorithms for computing $I[X;Y]$ with the KSG estimator\cite{jidt}.
The first is more accurate for small sample sizes but more biased; the
second is more accurate for larger sample sizes.  We use the second of
the two in this paper, as we have fairly long time series.  Our
algorithm sets $r_x$ and $r_y$ to the $x$ and $y$ distances to the
$k^{th}$ nearest neighbor.  One then counts the number of neighbors
within and on the boundaries of these kernels in each marginal space,
calling these sums $n_x$ and $n_y$, and finally calculates
\begin{align*}
  I[X;Y] = \psi(k) - \frac{1}{k}-\langle \psi(n_x) +\psi(n_y) \rangle + \psi(n)
  ~,
\end{align*}
where $\psi$ is the digamma function\footnote{The formula for the
  other KSG estimation algorithm is subtly different; it sets $r_x$
  and $r_y$ to the maxima of the $x$ and $y$ distances to the $k$
  nearest neighbors.}.  This estimator has been demonstrated to be
robust to variations in $k$ as long as $k\ge4$\cite{jidt}.

In this paper, we employ the Java Information Dynamics Toolkit (JIDT)
implementation of the KSG estimator\cite{jidt}.  The computational
complexity of this implementation is $\mathcal{O}(kN\log N)$, where
$N$ is the length of the time series and $k$ is the number of
neighbors being used in the estimate.  While this is more expensive
than traditional binning $(\mathcal{O}(N)$), it is bias corrected,
allows for adaptive kernel bandwidth to adjust for under- and
over-sampled regions of space, and is both model and parameter free
(aside from $k$, to which it is very robust).

\section{Applying \mytau to select reconstruction parameters }
\label{sec:results}

In this section, we demonstrate how to use \mytau to choose parameter
values for delay-reconstruction forecast models.  We do this for
several synthetic examples, as well as for sensor data from several
laboratory experiments.  For the discussion that follows, we use the
term ``\mytau-optimal'' to refer to the parameter values ($m$ and
$\tau$)
%$$ [m_{MASE},\tau_{MASE}] = \min_{m\in\{1,\dots,15\};\tau\in \{1,\dots,50\}}MASE(m,\tau,X_j,F)$$
that provided the best match between the forecast and the true
continuation.  
%
% In all cases, these \mytau-optimal values were identical to those
% found via exhaustive search over a wide range of $m$ and $\tau$.

To evaluate a forecast model, we divide the signal into two parts: the
initial training signal $\{x_j\}^n_{j=1}$---the first $n$ elements of
the time series---and the test signal $\{c_\ell\}^{k+n+1}_{\ell=n+1}$,
where $k$ is the length of the prediction.  We build a delay
reconstruction from the $x_j$ (i.e., a sequence of points
$[x_{j},x_{j-\tau},\dots,x_{j-(m-1)\tau)}]^T$), use it to generate a
prediction $\{\hat{x}_\ell\}^{k+n+1}_{\ell=n+1}$, and then use the
Mean Absolute Scaled Error\cite{MASE} to compare the prediction to the
test signal:
% over a range of $m$ and $\tau$  [[said just above]]

%
\begin{align*}
  MASE = \sum_{\ell=n+1}^{k+n+1} \frac{|\hat{x}_\ell-c_\ell|}{\frac{k}{n-1}\sum^n_{j=2}|x_{j}-x_{j-1}|}
\end{align*}
$MASE$ is a normalized measure: the scaling term in the denominator is
the average in-sample forecast error for a random-walk
prediction---which uses the previous value in the observed signal as
the forecast---calculated over the training signal.  That is, $MASE<1$
means that the prediction error in question was, on the average,
smaller than the in-sample error of a random-walk forecast on the
training portion of the same data.  Analogously, $MASE>1$ means that
the corresponding prediction method did \emph{worse}, on average, than
the random-walk method. 

While its comparative nature may seem odd, this error metric allows
for fair comparison across varying methods, prediction horizons, and
signal scales, making it a standard error measure in the forecasting
literature---and a good choice for the study described in the
following sections, which involve a number of very different signals.

\subsection{Synthetic examples}
\label{subsec:synthetic}

In this Section, we apply \mytau to some standard synthetic examples,
both maps (H\'{e}non, logistic) and flows: the classic Lorenz
system\cite{lorenz} and the more-recent ``Lorenz 96'' atmospheric
model\cite{lorenz96Model}.  We construct the traces for the Lorenz
experiments using a standard fourth-order Runge-Kutta solver on the
associated differential equations, with a timestep of $\frac{1}{64}$,
for 60,000 time steps.  For the maps, we simply iterate the difference
equations 60,000 times.  In all cases, we discard the first 10,000
points of each trajectory to remove transient behavior, then sample
individual state variables to produce different scalar time-series
data sets.  We reconstruct the dynamics from those traces using
different values of the dimension $m$ and delay $\tau$ and compute
\mytau for each of those reconstructed trajectories.  We then use
Lorenz's classic method of analogues (LMA) \cite{lorenz-analogues} to
generate forecasts of each trace, compute their $MASE$ scores as
described above, and discuss their relationships to the \mytau values
for the corresponding time series.  For simplicity, in this initial
discussion we perform a series of one-step-ahead predictions,
rebuilding the model at each step.  For the \mytau calculations, this
means that we estimate $I[\mathcal{S}_j,X_{j+1}]$, with
$\mathcal{S}_j= [X_{j},X_{j-\tau},\dots,X_{j-(m-1)\tau)}]^T$.
% for a range of $m=2,\dots,15$ and $\tau=1,\dots,50$.  
In Section~\ref{subsec:predictionhorizon} we expand this discussion by
increasing the prediction horizon; in Section~\ref{subsec:datalength},
we consider the effects of the length of the traces.

\subsubsection{Flow examples}
\label{subsubsec:lorenz96}

The Lorenz 96 system\cite{lorenz96Model} is defined by a set of $K$
differential equations in the state variables $\xi_1\dots\xi_K$:
\begin{equation}\label{eq:lorenz96}
  \dot{\xi}_k= (\xi_{k+1}-\xi_{k-2})(\xi_{k-1})-\xi_k + F \nonumber
\end{equation}
for $k=1,\dots,K$, where $F\in \mathbb{R}$ is a constant forcing term
that is independent of $k$. In the following discussion we focus on
two parameter sets, $\{K=22,F=5\}$ and $\{K=47,F=5\}$, which produce
low- and high-dimensional chaos, respectively.  See \cite{KarimiL96}
for an explanation of this model and the associated parameters.

\begin{figure}[ht!]
        \centering
        \begin{subfigure}[b]{\columnwidth}
                \includegraphics[width=0.9\columnwidth]{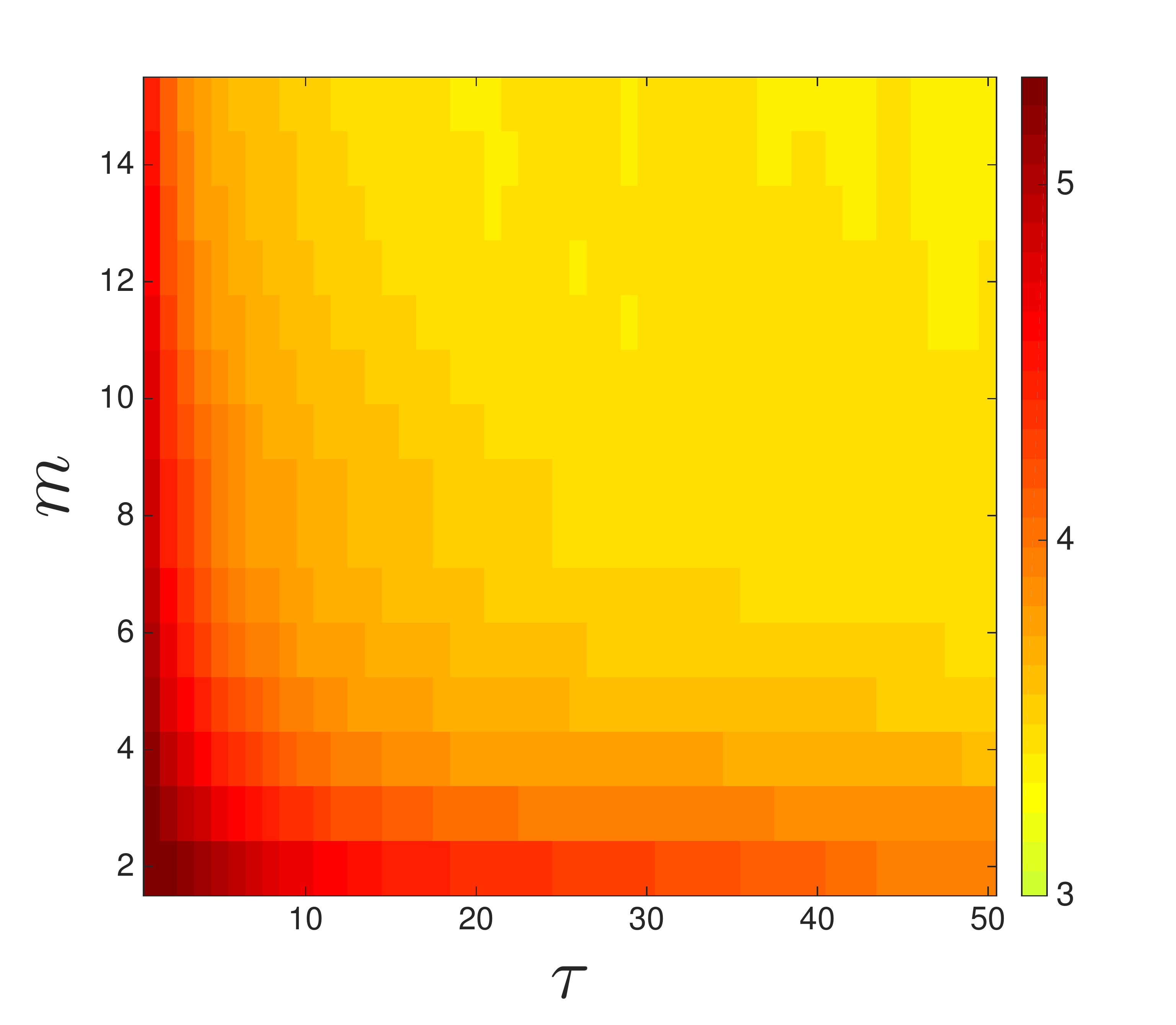}
                \caption{\mytau values for different delay
                  reconstructions of a representative trace from the
                  Lorenz 96 system with $\{K=22,F=5\}$.}
                \label{fig:L96N22F5SPI}
        \end{subfigure}%
        ~ %add desired spacing between images, e. g. ~, \quad, \qquad etc.

        \begin{subfigure}[b]{\columnwidth}
                \includegraphics[width=0.9\columnwidth]{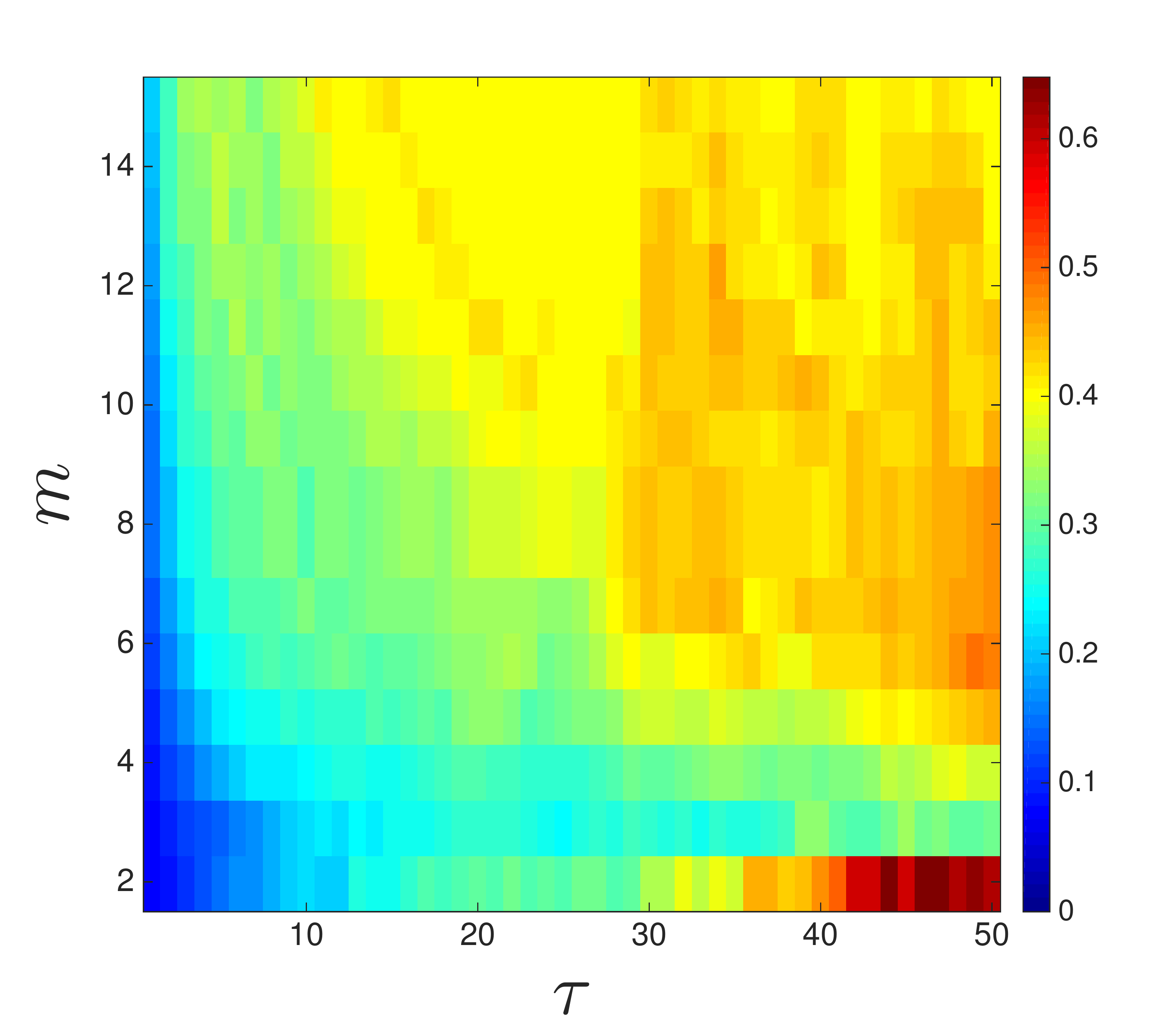}
                \caption{MASE scores for LMA forecasts on different
                  delay reconstructions of a representative trace of
                  the Lorenz 96 system with $\{K=22,F=5\}$. }
                \label{fig:L96N22F5MASE}
        \end{subfigure}

        %\caption{(b) Minimum MASE occurs at 0.0737 with $m=2$ and $\tau=1$.}\label{fig:tauandmL22}
                \caption{The effects of reconstruction parameter
                  values on \mytau and forecast accuracy for the
                  Lorenz 96 system}\label{fig:tauandmL96}
\end{figure}

Figure~\ref{fig:L96N22F5SPI} shows a heatmap of the \mytau values for
reconstructions of a representative trajectory from this system with
$\{K=22,F=5\}$, for a range of $m$ and $\tau$.  
% (The others were indistinguishable.)
Not surprisingly, this image reveals a strong dependency between the
values of the reconstruction parameters and the reduction in
uncertainty about the near future that is provided by the
reconstruction.  Very low $\tau$ values, for instance, produce delay
vectors with highly redundant coordinates, but which provide substantial
information about the future.  As mentioned in the first section of
this paper, standard heuristics only focus on minimizing redundancy between coordinates and choose the
$\tau$ value that minimizes the mutual information between the first
two coordinates in the delay vector.  For this trajectory, the
approach of Fraser \& Swinney\cite{fraser-swinney} yields $\tau=26$,
while standard dimension-estimation heuristics \cite{KBA92} suggest
$m=8$.  The \mytau value for a delay reconstruction built with those
parameter values is 3.463.  This is \emph{not}, however, the
\mytau-optimal reconstruction; choosing $m=2$ and $\tau=1$, for
instance, results in a higher value ($\mytau=5.303$)---i.e.,
signficantly more reduction in uncertainty about the future.  This may
be somewhat counter-intuitive, since each of the delay vectors in the
\mytau-optimal reconstruction spans far less of the data set and thus
one would expect points in that space to contain \emph{less}
information about the future.  Figure~\ref{fig:L96N22F5SPI} suggests,
however, that this in fact not the case; rather, that uncertainty
\emph{increases} with both dimension and time delay.
\label{page:increase-with-tau}

The question at issue in this paper is whether that reduction in
uncertainty about the future correlates with improved accuracy of an
LMA forecast built from that reconstruction.  Since the \mytau-optimal
choices
% 
% ---$m=2$ and $\tau = 1$, in Figure~\ref{fig:L96N22F5SPI}---
% 
maximize the shared information between the state estimator and
$X_{j+1}$, one would expect a delay reconstruction model built with
those choices to afford LMA the best leverage.  To test that
conjecture, we performed an exhaustive search with $m=2,\dots,15$ and
$\tau=1,\dots,50$.  For each $\{m, \tau\}$ pair, we used LMA to
generate forecasts from the corresponding reconstruction, computed
their $MASE$ scores, and plotted the results in a heatmap similar to
the one in Figure~\ref{fig:L96N22F5SPI}.  As one would expect, the
$MASE$ and \mytau heatmaps are generally antisymmetric.  This
antisymmetry breaks down somewhat for low $m$ and high $\tau$, where
the forecast accuracy is low even though the reconstruction contains a
lot of information about the future.
%  Notice that in some of the regions of the heatmap we are not
% interested in, i.e., away from max/min respectively, some of the
% symmetry breaks down. In particular, for $m=2$ and $\tau>40$ the
% MASE scores seem to be worse than one would expect given the
% corresponding \mytau values.  What we suspect is occuring in this
% region is that the attractor is overfolded by using too large of a
% $\tau$ and too low of an embedding dimension.  In this case, the
% over folding combined by the projection of they dynamics incurred by
% the projection are causing LMA to do poorly due to the increase in
% false crossings in the reconstruction. That is, while there is still
% a fair amount of information about the future, an increase in false
% crossings in the dynamics confuse the LMA model as to which forward
% path to take. Notice, this effect is mitigated by adding another
% dimension. This may suggest that \mytau should only be fully trusted
% in the region of interest---near maximum in the \mytau surface and
% equivalently near minimum in the MASE surface.
We suspect that this is due to a combination of overfolding (due to
too-large values of $\tau$) and projection (low $m$).  Even though
each point in such a reconstuction may contain a lot of information
about the future, the false crossings created by this combination of
effects pose problems for a near-neighbor forecast strategy like LMA.
The improvement that occurs if one adds another dimension is
consistent with this explanation.  Notice, too, that this effect only
occurs far from the maximum in the \mytau surface---the area that is
of interest if one is using \mytau to choose parameter values for
reconstruction models.

In general, though, maximizing the redundancy between the state
estimator and the future does appear to minimize the resulting
forecast error of LMA.  Indeed, the maximum on the surface of
Figure~\ref{fig:L96N22F5MASE} ($m=2,\tau=1$) is exactly the minimum on
the surface of Figure~\ref{fig:L96N22F5SPI}.  The accuracy of this
forecast is more than five times higher ($MASE=0.0737$) than that of a
forecast constructed with the parameter values suggested by the
standard heuristics ($0.3787$).  Note that the optima of these
surfaces may be broad: i.e., there may be \emph{ranges} of $m$ and
$\tau$ for which \mytau and $MASE$ are optimal, and roughly constant.
In these cases, it makes sense to choose the lowest $m$ on the
plateau, since that minimizes computational effort, data requirements,
and noise effects; see \cite{joshua-pnp} for a full discussion of
this.

\begin{figure}
        \centering
        \begin{subfigure}[b]{\columnwidth}
                \includegraphics[width=0.9\columnwidth]{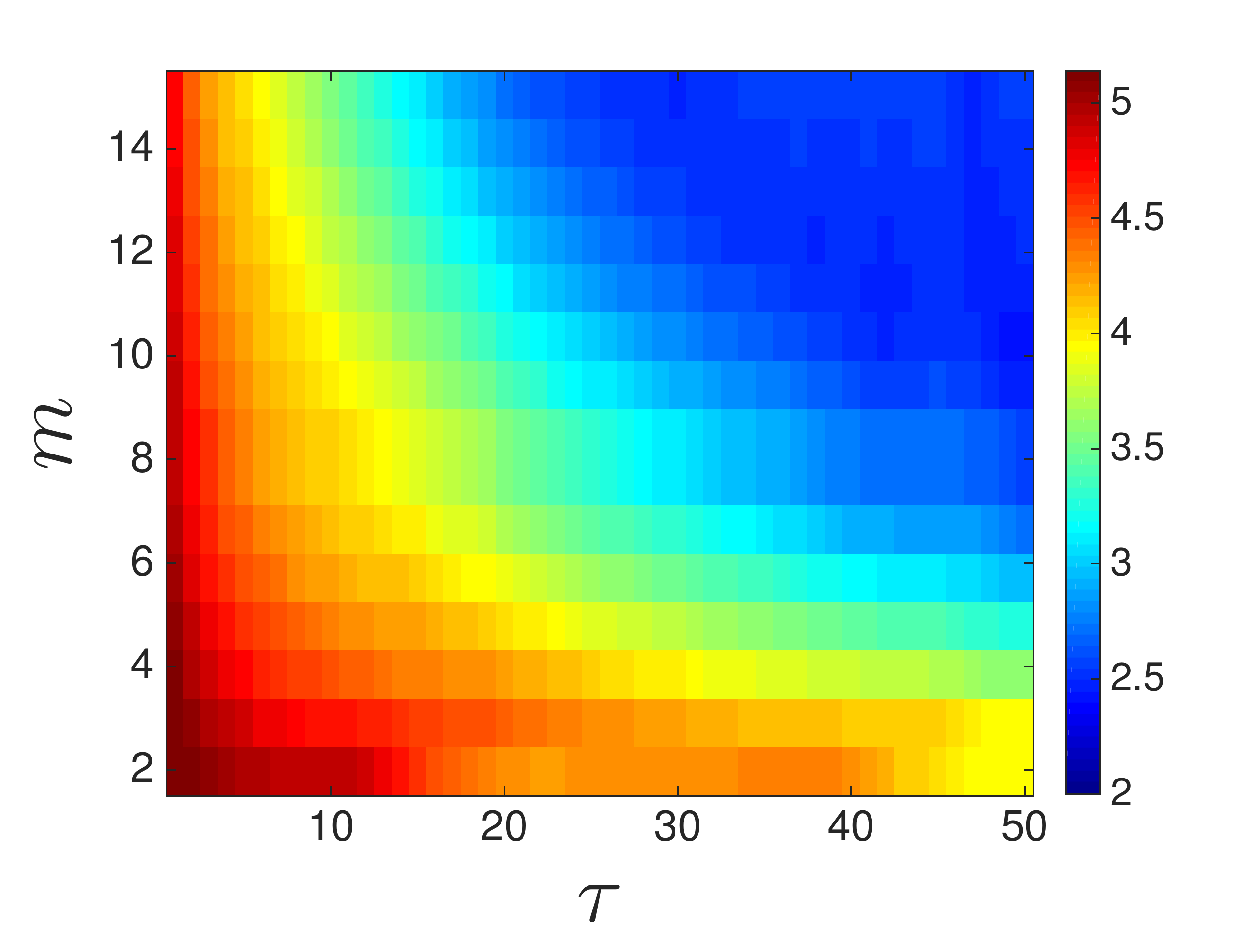}
                \caption{\mytau values for different delay
                  reconstructions of a representative trace from the
                  Lorenz 63 system.}
                \label{fig:L63SPI}
        \end{subfigure}%
        ~ %add desired spacing between images, e. g. ~, \quad, \qquad etc.

        \begin{subfigure}[b]{\columnwidth}
                \includegraphics[width=0.9\columnwidth]{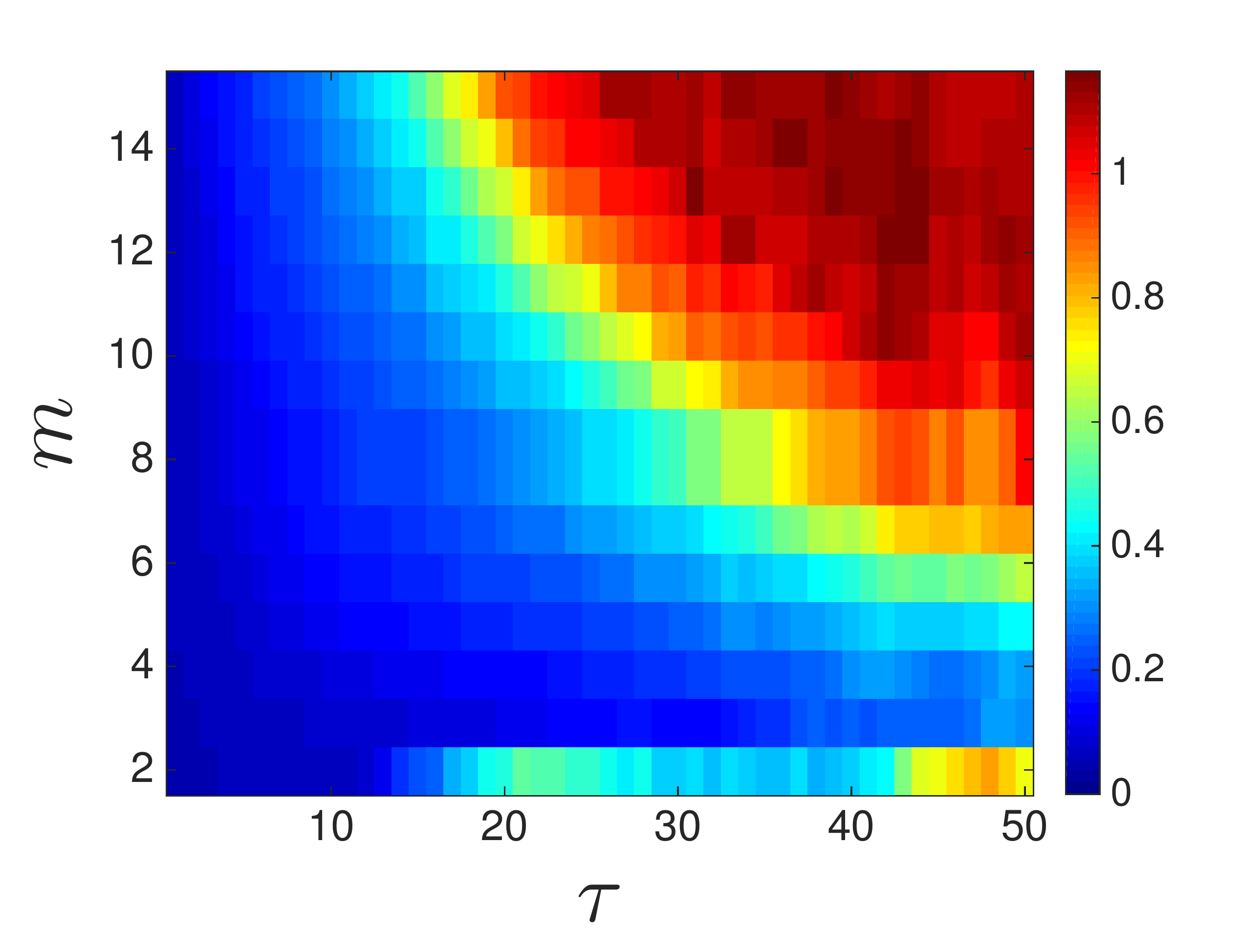}
                \caption{$MASE$ scores for LMA forecasts on different
                  delay reconstructions of a representative trace of
                  the Lorenz 63 system. }
                \label{fig:L63MASE}
        \end{subfigure}

        %\caption{(b) Minimum MASE occurs at 0.0737 with $m=2$ and $\tau=1$.}\label{fig:tauandmL22}
                \caption{The effects of reconstruction parameter
                  values on \mytau and forecast accuracy for the
                  Lorenz 63 system}\label{fig:tauandmL63}
\end{figure}

While the results discussed in the previous paragraph do provide a
preliminary validation of the claim that one can use \mytau to select
good parameter values for delay reconstruction-based forecast
strategies, they only involve a single example system.  Similar
experiments on traces from the Lorenz 96 system with different
parameter values $\{K=47,F=5\}$ show identical results---indeed, the
heatmaps are visually indistinguishable from the ones in
Figure~\ref{fig:tauandmL96}.  Figure~\ref{fig:tauandmL63} shows
heatmaps of \mytau and $MASE$ for similar experiments on the classic
``Lorenz 63'' system\cite{lorenz}:
\begin{eqnarray*}
  \dot{x} &=& \sigma(y-x)\\
  \dot{y} &=& x(\rho -z)-y \\
  \dot{z} &=& xy -\beta z
\end{eqnarray*}
with the typical chaotic parameter selections: $\rho=28, \sigma=10$,
and $\beta=8/3$.
As in the Lorenz 96 case, the heatmaps are generally antisymmetric,
confirming that maximizing \mytau is roughly equivalent to minimizing
$MASE$.  Again, though, the antisymmetry is not perfect; for high
$\tau$ and low $m$, the effects of projecting an overfolded attractor
cause false crossings that trip up LMA.  As before, adding a dimension
mitigates this effect by removing these false crossings.  Both the
Lorenz 63 and Lorenz 96 plots show a general decrease in
predictability for large $m$ and high $\tau$, with roughly hyperbolic
equipotentials dividing the colored regions\footnote{Note that the
  color map scales are not identical across all heatmap figures in
  this paper; rather, they are chosen individually, to bring out the
  details of the structure of each experiment.}.  The locations and
heights of these equipotentials differs because the two signals are
not equally easy to predict.  This matter is discussed further at the
end of this section.

% : notice for example in the $m=2$ case we again see increased MASE for
% $\tau>20$, in this case the wings have folded back on themselves and
% many false crossings have been introduced in this 2D projection. This
% is mitigated by adding another dimension for these folded wings to
% exist in. This is seen as well in Fig.\ref{fig:tauandmL63} by there
% being higher \mytau for a longer range of $\tau$ at $m=3$.

Numerical \mytau and $MASE$ values for LMA forecasts on different
reconstructions of both Lorenz systems are tabulated in the top three
rows of Table~\ref{tab:myTauParams}, along with the reconstruction
parameter values that produced those results.
\begin{table*}[tb!]
  \caption{$MASE$ values for various delay reconstructions of the
    different examples studied here.  $MASE_H$ is the representative
    accuracy of LMA forecasts that use delay reconstructions with
    parameter values ($m_{\mytau}$ and $\tau_{\mytau}$) chosen via
    standard heuristics for the corresponding traces---the methods of
    false neighbors \cite{KBA92} and time-delayed mutual information
    \cite{fraser-swinney}, respectively.  Similarly, $MASE_{\mytau}$
    is the accuracy of LMA forecasts that use reconstructions built
    with the $m$ and $\tau$ values that maximize \mytau, and $MASE_E$
    is the error of the best forecasts for each case, found via
    exhaustive search over the $m,\tau$ parameter
    space. \alert{${**}$}: on these signals the standard heuristics
    failed.}
  \begin{center}
    \begin{tabular}{lccccccccc}
      \hline\hline Signal & $MASE_H$ & $\tau_H$& $m_H$ & $MASE_{\mytau}$ & $\tau_{\mytau}$ & $m_{\mytau}$ & $MASE_E$ & $\tau_E$& $m_E$\\
      \hline
      Lorenz-96 $K=22$ &0.3787&26&8&0.0737&1& 2& 0.0737 & 1& 2\\
      Lorenz-96 $K=47$ &1.007&31&10&0.1156& 1& 2&0.1156& 1& 2\\
      % \col           & 2  &2 & 2&15\\%\\& $0.599  \pm 0.211 $ & $0.571\pm0.002$&  $0.513$\\%\pm 0.003$ \\
      %\gcc           & 10&2 &2 &15\\%& $1.837 \pm0.016 $ & $0.951 \pm 0.001$ & $0.943$\\% \pm 0.001$ \\
      %SFI A & 2& 1&1& 1\\
      Lorenz 63 &0.2215&12&5& 0.0509&1&3& 0.0506&1&2\\
      % R\"ossler \\
      \hline
      H\'enon Map & \alert{$^{**}$}& \alert{$^{**}$}&\alert{$^{**}$} &3.814e-04&1&2 &3.814e-04&1 &2 \\
      Logistic Map & \alert{$^{**}$} &\alert{$^{**}$}&\alert{$^{**}$} &1.680e-05&1& 1&1.680e-05& 1&1 \\
      \hline\hline
    \end{tabular}
  \end{center}
  \label{tab:myTauParams}
\end{table*}%
The data in this table bring out two important points.  First, as
suggested by the heatmaps, the $m$ and $\tau$ values that maximize
\mytau (termed $m_{\mytau}$ and $\tau_{\mytau}$ in the table legend)
are close, or identical, to the values that minimize $MASE$ ($m_E$ and
$\tau_E$) for all three Lorenz systems.  This is notable because---as
discussed in Section~\ref{subsec:datalength}---the former can be
estimated quite reliably from a small sample of the trajectory in only
a few seconds of compute time, whereas the exhaustive search that is
involved in computing $m_E$ and $\tau_E$ for
Table~\ref{tab:myTauParams} required close to 30 hours of CPU time per signal.
% were estimated from 750 calculations using $\approx 45,000$ data
% points and took 6.143 minutes---and can be calculated reliably from
% even less data---750 points per calculation, which reduced computing
% time to 8.5904 seconds, as discussed in Section~\ref{sec:where}---and
% the latter were computed via exhaustive search across 750 predictions
% of 5,000 points using a training set of 45,000 points, requiring
% 29.4467 hours.
A second important point that is apparent from the Table is that
delay reconstructions built using the traditional heuristics---the
values with the $H$ subscript---were comparatively ineffective for the
purposes of LMA-based forecasting.  This is notable because that is
the default approach in the literature on state-space based
forecasting methods for dynamical systems.

% As was the case with Lorenz 96 in the previous discussion choosing
% reconstruction parameters using the standard heuristics
% ($m=5,\tau=12$) resulted in significantly less \mytau in the
% associated LMA model (3.348 versus the maximum of 4.5520) and worse
% forecast accuracy MASE$_{H}=0.2215$ as opposed to
% MASE$_{\mytau}=0.0471$---a full order of magnitude increase in
% accuracy! Something interesting to point out in this is that
% maximizing \mytau did not find the optimal minimization of forecast
% error, but it was close. MASE$_{E}=0.04382$ and was associated with a
% \mytau$ =4.518$. The maximal $\mytau$ was at 4.552 and resulted in a
% slightly higher error of MASE$_{\mytau}=0.0471$. This small difference
% in \mytau and MASE show that maximizing \mytau may only suggest
% near-optimal reconstruction parameters due to statistical
% fluctuation. However, note that maximizing \mytau was significantly
% better than using standard heuristics.

\begin{figure}[ht!]
        \centering
        \begin{subfigure}[b]{\columnwidth}
                \includegraphics[width=0.9\columnwidth]{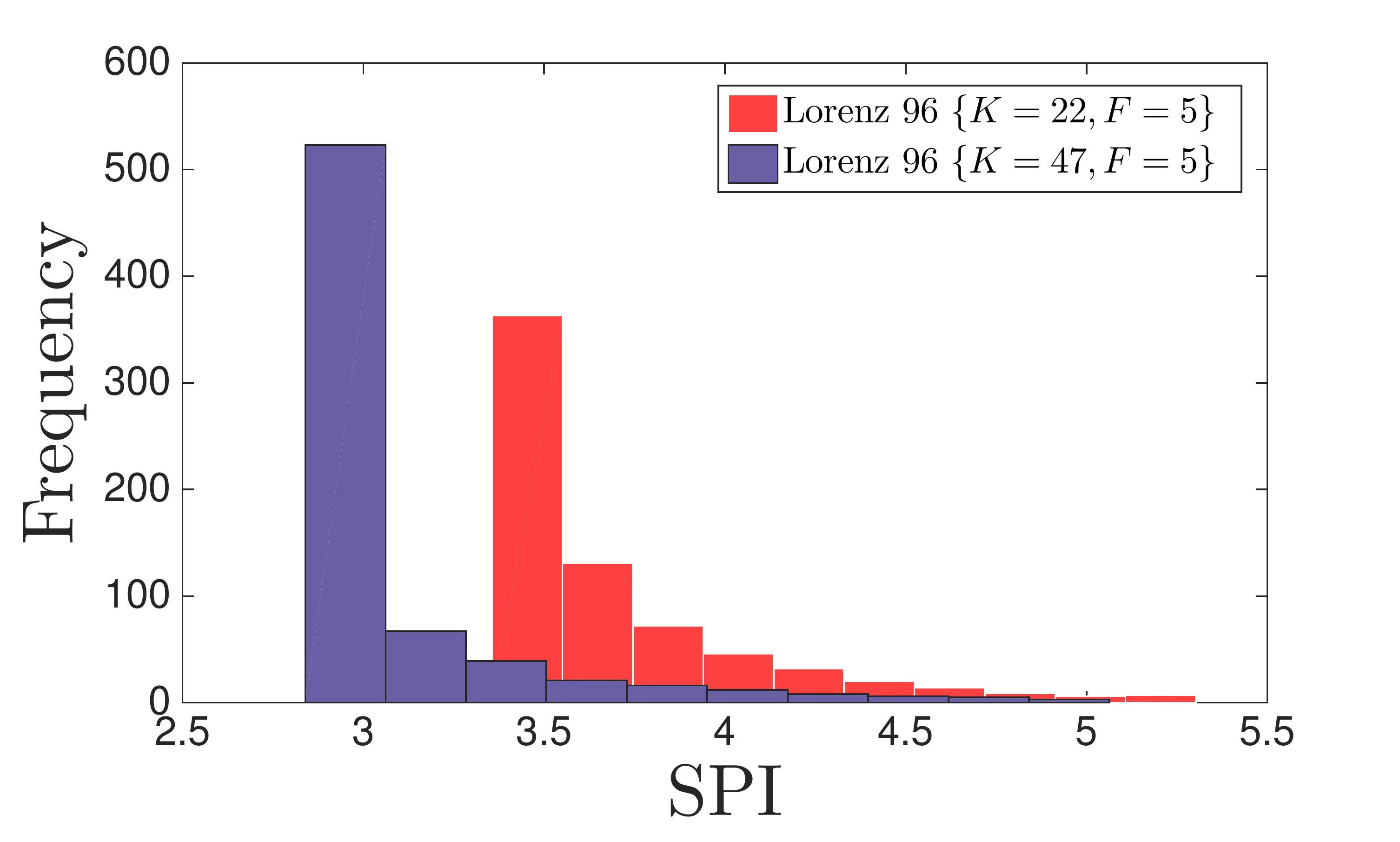}
                \caption{\mytau}
                \label{fig:L96SPIHIST}
        \end{subfigure}%
        ~ %add desired spacing between images, e. g. ~, \quad, \qquad etc.

        \begin{subfigure}[b]{\columnwidth}
                \includegraphics[width=0.9\columnwidth]{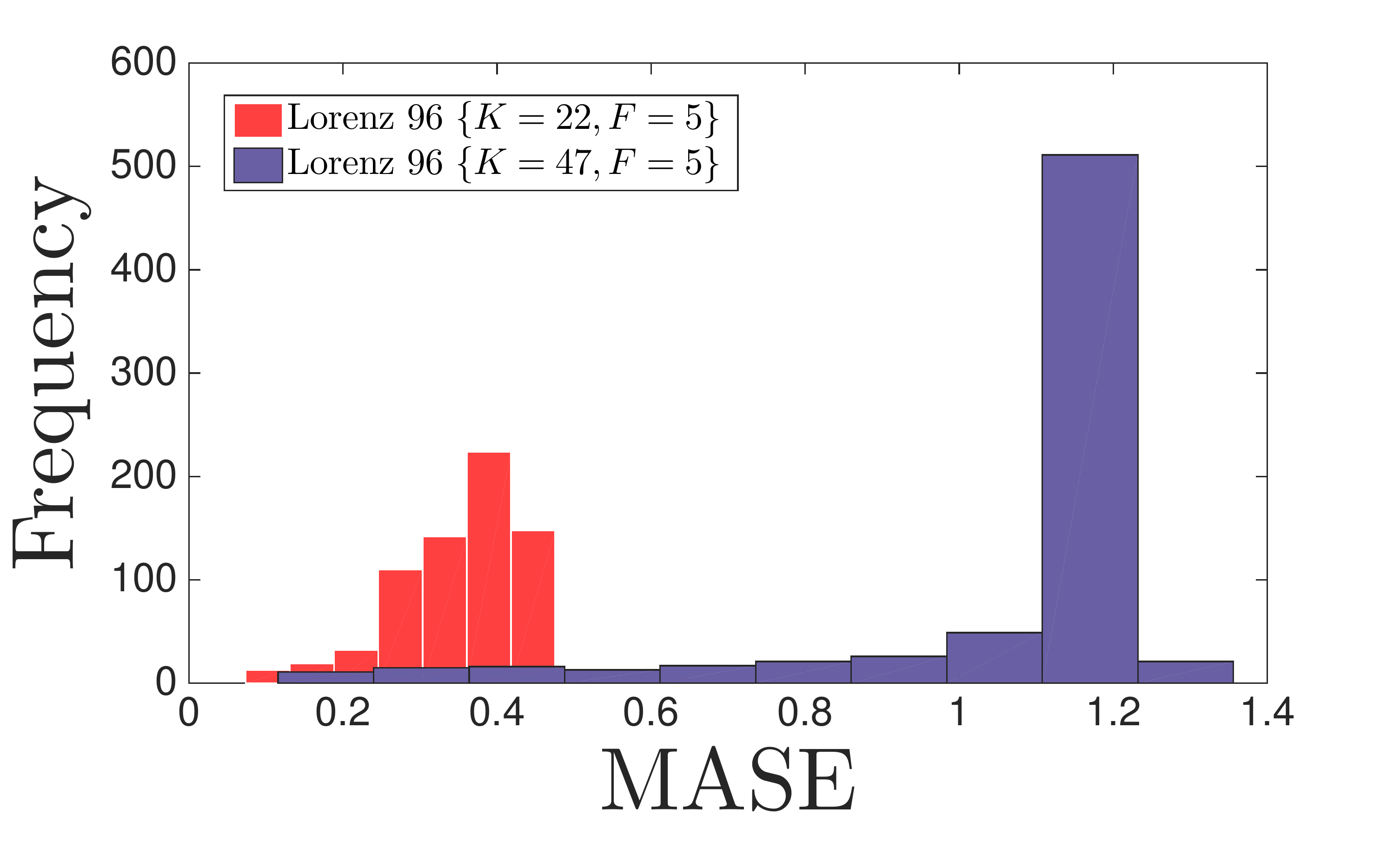}
                \caption{$MASE$}
                \label{fig:L96MASEHIST}
        \end{subfigure}

        %\caption{(b) Minimum MASE occurs at 0.0737 with $m=2$ and $\tau=1$.}\label{fig:tauandmL22}
                \caption{Histograms of \mytau and $MASE$ values for
                  representative traces from the Lorenz 96
                  $\{K=22,F=5\}$ and $\{K=47,F=5\}$ systems for all
                  $\{m,\tau\}$ values in Figures~\ref{fig:tauandmL96}
                  and~\ref{fig:tauandmL63}.}
\label{fig:L96HistCompares}
\end{figure}

A close comparison of Figures~\ref{fig:tauandmL96}
and~\ref{fig:tauandmL63} brings up another important point: some time
series are harder to forecast than others.
Figure~\ref{fig:L96HistCompares} breaks down the details of the two
suites of Lorenz-96 experiments, showing the distribution of \mytau
and $MASE$ values for all of the reconstructions.
Although there is some overlap in the $K=47$ and $K=22$
histograms---i.e., best-case forecasts of the former are better than
most of the forecasts of the latter---the $K=47$ traces generally
contain less information about the future and thus are harder to
forecast accurately.  

%\subsubsection{Lorenz 63 \alert{ and R\"ossler}}
% \subsubsection{Lorenz 63}

%\alert{Maybe add discussion with R\"ossler}

\subsubsection{Map examples}

Delay reconstruction of discrete-time dynamical systems, while
possible in theory, can be problematic in practice.  Although the
embedding theorems do apply in these cases, the heuristics for
estimating $m$ and $\tau$ often fail.  The time-delayed mutual
information of \cite{fraser-swinney}, for example, may decay
exponentially, without showing any clear minimum.  And the lack of
spatial continuity of the orbit of a map violates the underlying idea
behind the method of \cite{KBA92}.  State space-based forecasting
methods can, however, be very useful in generating predictions of
trajectories from systems like this---\emph{if} one has a
reconstruction that is faithful to the true dynamics.

In view of this, it would be particularly useful if one could use
\mytau to choose embedding parameter values for maps.  This section
explores that notion using two canonical examples, shown in the bottom
two rows of Table~\ref{tab:myTauParams}.  For the H\'enon map,
\begin{eqnarray}
  x_{n+1} &=& 1-ax^2_n +y_n \nonumber \\
  y_{n+1} &=& bx_n \nonumber
\end{eqnarray}
with $a=1.4$ and $b=0.3$, the \mytau-optimal parameter values were
$m=2$ and $\tau=1$.  As in the flow examples, these were identical to
the values that minimized $MASE$.  These parameter values make sense, of course;
a first-return map of the $x$ coordinate is effectively the H\'enon
map, so $[x_j,x_{j-1}]$ is a perfect state estimator (up to a scaling
term).  But in practice, of course, one rarely knows the underlying
dynamics of the system that generated a time series, so the fact that
one can choose good reconstruction parameter values by maximizing
\mytau is notable---especially since standard heuristics for that
purpose fail in this system.

The same pattern holds for the logistic map, $x_{n+1} = rx_n(1-x_n)$,
with $r=3.65$: the \mytau-optimal parameter values coincide
with the minimum of the $MASE$ surface.  As in the H\'{e}non example,
these values ($m=1$ and $\tau=1$) make complete sense, given the form
of the map.  But again, one does not always know the form of the
system that generated a given time series.  In the case of the
logistic map, the standard heuristics fail, but \mytau clearly
indicates that one does not actually need to reconstruct these
dynamics---rather, near-neighbor forecasting \emph{on the time series
  itself} is the best approach.

% Neither of these parameter findings are interesting in and of
% themselves but this shows a unique feature of \mytau. It can be
% applied to select optimal parameters for forecasting, \emph{even if
%   the system being observed is a map}---something that stymies all the
% traditional heuristics. It could be argued that selecting these
% parameters is obvious from the structure of the system, and this is
% true. However, in the real-world a practitioner is rarely afforded the
% luxury of knowing this structure. Moreover, maps are commonly observed
% in practice due to finite sensing and standard heuristics cannot be
% applied for the above mentioned reasons. However, \mytau provides a
% work around to perform parameter selection in this difficult,
% important, and often neglected case.

% In Tables~\ref{tab:myTauError} and~\ref{tab:myTauParams} we tabulate
% all parameters and errors discussed in this section. In the synthetic
% case it seems that maximizing \mytau minimizes forecast accuracy of
% LMA. In the next section, we turn our attention to actual experimental
% time series.

\subsection{Selecting reconstruction parameters of experimental time series}
\label{subsec:experimental}

The results in the previous section provide a preliminary verification
of the conjecture that maximizing \mytau minimizes forecast accuracy
of LMA, for both maps and flows.  While experiments with synthetic
examples are useful, they do not call the really important aspect of
that research question: whether \mytau is a useful way to choose
parameter values for delay reconstruction-based forecasting of
real-world data, where the time series are noisy and perhaps short,
and one does not know the dimension of the underlying system---let
alone its governing equations.  In this section, we turn our attention
to that question using experimental data from two different dynamical
systems: a far-infrared laser and a laboratory computer-performance
experiment.

\subsubsection{A Far-Infrared Laser}
\label{subsubsec:laser}

\begin{figure}[ht!]
        \centering
        \begin{subfigure}[b]{\columnwidth}
                \includegraphics[width=\columnwidth]{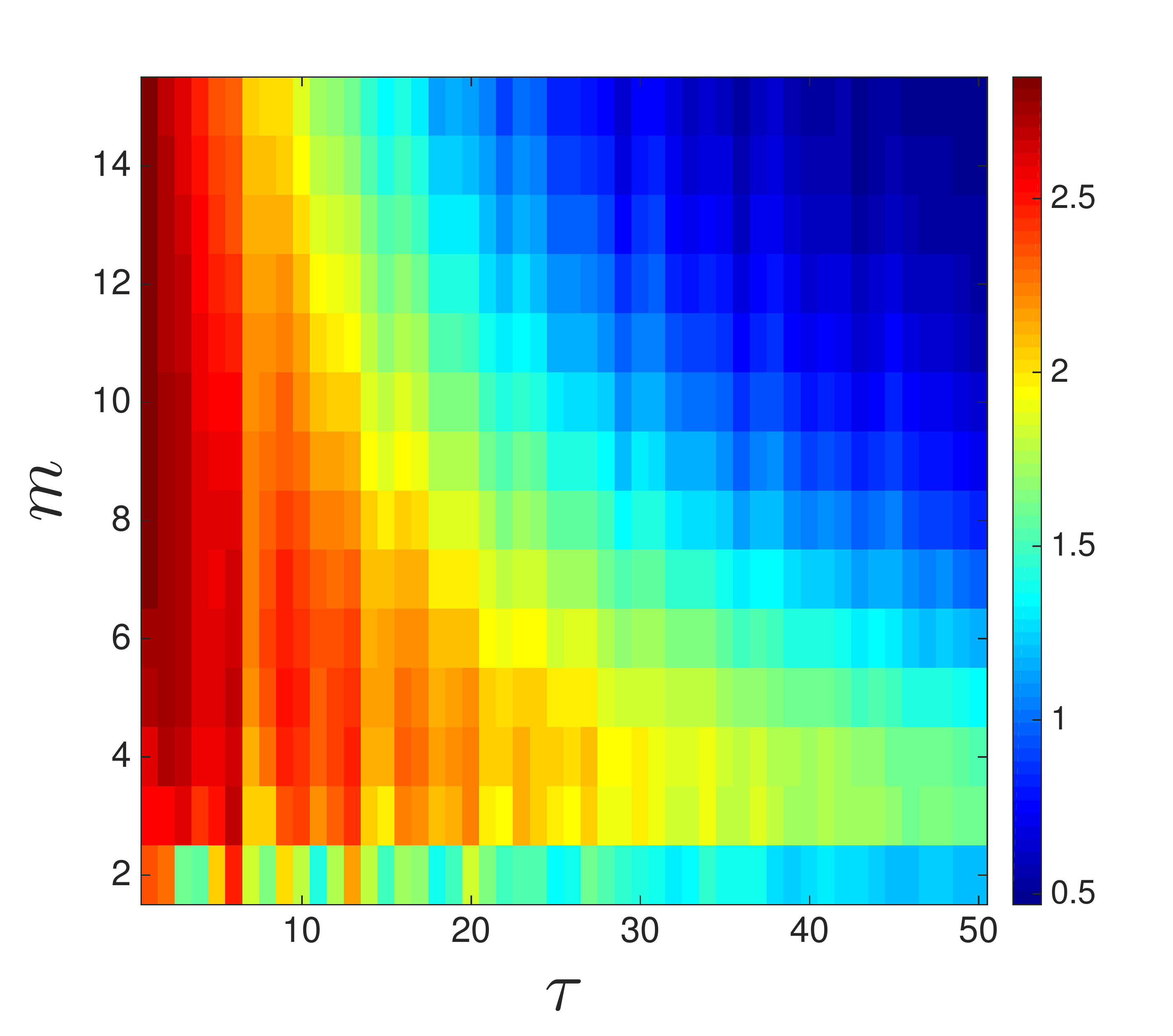}
                \caption{\mytau values for different delay
                  reconstructions of SFI Dataset A.  }
                \label{fig:LASERMASE}
        \end{subfigure}%
        ~ %add desired spacing between images, e. g. ~, \quad, \qquad etc.
%          %(or a blank line to force the subfigure onto a new line)

        \begin{subfigure}[b]{\columnwidth}
                \includegraphics[width=\columnwidth]{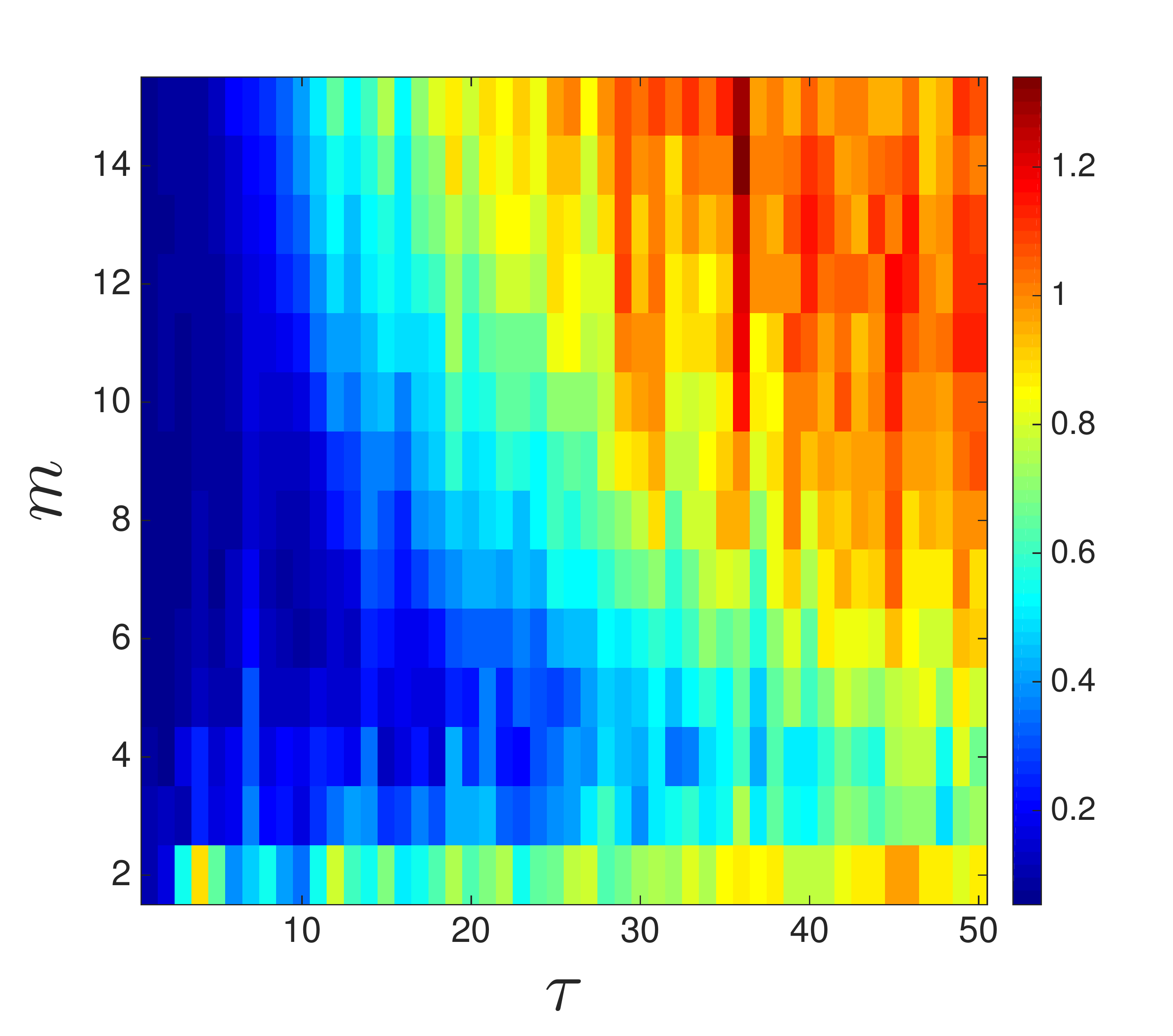}
                \caption{MASE scores for LMA forecasts on different
                  delay reconstructions of SFI Dataset A.}
                \label{fig:LASERSPI}
        \end{subfigure}
                \caption{The effects of reconstruction parameter
                  values on \mytau and forecast accuracy for ``Dataset
                  A'' from the Santa Fe Institute time-series
                  prediction competition.}
\label{fig:MASEandmyTAULaser}
\end{figure}

A canonical test case in the forecasting literature is the so-called
``Dataset A" from the Santa Fe Institute prediction
competition\cite{weigend-book}, which was gathered from a far-infrared
laser.  As in the synthetic examples in the previous section, the
\mytau and $MASE$ heatmaps (Figure~\ref{fig:MASEandmyTAULaser}) are
largely antisymmetric for this signal.
Again, there is a band across the bottom of each image because of the
combined effects of overfolding and projection.  Note the resemblance
between Figures ~\ref{fig:MASEandmyTAULaser} and
~\ref{fig:tauandmL63}: the latter resemble ``smoothed'' versions of
the former.  It is well known\cite{weigend-book} that the SFI A
dataset is well described by the Lorenz 63 system with some added
noise, so this similarity is both unsurprising and reassuring.  LMA
forecasts using the \mytau-optimal reconstruction of this trace were
more accurate than similar forecasts using a reconstruction built
using traditional heuristics ($MASE_{\mytau}= 0.0592$ versus
$MASE_H=0.0733$) and only slightly worse than the optimal value
($MASE_E=0.0538$).
%  $SPI=2.847, 2.734$ and $2.814$, respectively.
%\alert{Josh, I know we talked about not including the $m$ and $\tau$
%  values in the \col example because of the broad-optimum issue, but I
%  think it would be useful to have them here---otherwise people may
%  wonder, since the correspondence between $m_{\mytau},\tau_{\mytau}$
%  and $m_E,\tau_E$ was such an important point in the previous section
%  (and the Table).  
%If they're not quite the same, we can mention the
%  broad-optimum thing here, and give a forward pointer to the
%  discussion in the following section.}  
However, the values of $\{m_{\mytau},\tau_{\mytau}\}$ and
$\{m_E,\tau_E\}$ are not identical for this signal.  This is because
the optima in the heatmaps in Figure~\ref{fig:MASEandmyTAULaser} are
bands, rather than unique points---as was the case in the synthetic
examples in Section~\ref{subsec:synthetic}.  In a situation like this,
a range of $\{m,\tau\}$ values are statistically indistinguishable,
from the standpoint of the forecast accurary afforded by the
corresponding reconstruction.  The values suggested by the \mytau
calculation ($m_{\mytau}=9$ and $\tau_{\mytau}=1$) and by the
exhaustive search ($m_E=7$, $\tau_E=1$) were all on this
plateau\footnote{The values suggested by the traditional heuristics,
  $m_H=7$ and $\tau_H=3$, were off the shoulder of that plateau.}.
Again, it appears that one can use \mytau to choose good parameter
values for delay reconstruction-based forecasting, but SFI A is only a
single trace from a fairly simple system.

\subsubsection{Computer Performance Dynamics}
\label{subsubsec:computer}

Laboratory experiments on computer performance dynamics have shown
that these high-dimensional nonlinear systems exhibit a range of
interesting deterministic dynamical
behaviors\cite{zach-IDA10,mytkowicz09}.  Both hardware and software
play roles in these dynamics; changing either one can cause
bifurcations from periodic orbits to low- and high-dimensional chaos.
This rich range of behavior makes computer performance dynamics an
ideal final test case for this paper.

%It also has important practical implications; these
%dynamics, which arise from the deterministic, nonlinear interactions
%between the hardware and the software, have profound effects on
%execution time and memory use.
Collecting observations of the performance of a running computer
requires some significant engineering.  Basically, one programs the
microprocessor's onboard hardware performance monitor to observe the
quantities of interest, then stops the program execution at
100,000-instruction intervals---the unit of time in these
experiments---and reads off the contents of those registers.
Interested readers can find a detailed description of this custom
measurement infrastructure in\cite{todd-phd,mytkowicz09}.  The signals
that are produced by this apparatus are scalar time-series
measurements of system metrics like processor efficiency ({\it e.g.,}
\ipc, which measures how many instructions are being executed, on the
average, in each clock cycle) or memory usage ({\it e.g.,} how often
the processor had to access the main memory during the measurement
interval).

Here, for conciseness, we focus on \emph{processor} performance traces
from two different programs, one simple and one complex, running on
the same Intel i7-based computer.  The first is four lines of C (\col)
that repeatedly initializes a $256 \times 256$ matrix in column-major
order.  The second is a much more complex program: the \gcc compiler
from the SPEC 2006CPU benchmark suite\cite{spec2006}.  The performance
traces of these two programs contained 147,925 points and 45,545
points, respectively.  Since computer performance dynamics result from
a composition of hardware and software, these two experiments involve
two different dynamical systems, even though the programs are running
on the same computer.  But since other effects could be at
work---housekeeping by the operating system, etc.---we repeated each
experiment 15 times for a total of 30 traces.  We have performed
similar forecast experiments using other processor and memory
performance metrics gathered during the execution of a variety of
programs on several different computers \cite{josh-pre}.  Our
preliminary analysis indicates that the results described in the rest
of this section hold for those traces as well.

%The dynamical differences are visually apparent from the traces in
%Figure~\ref{fig:computerdata}; they are mathematically apparent from
%nonlinear time-series analysis of embeddings of those
%data\cite{mytkowicz09}, as well as in calculations of the information
%content of the two signals.  Among other things, \gcc has much less
%predictive structure than \col and is thus much harder to
%forecast\cite{josh-pre}.  These attributes make this a useful pair of
%experiments for an exploration of the utility of reduced-order
%forecasting.

\begin{figure}[ht!]
        \centering
        \begin{subfigure}[b]{\columnwidth}
                \includegraphics[width=0.9\columnwidth]{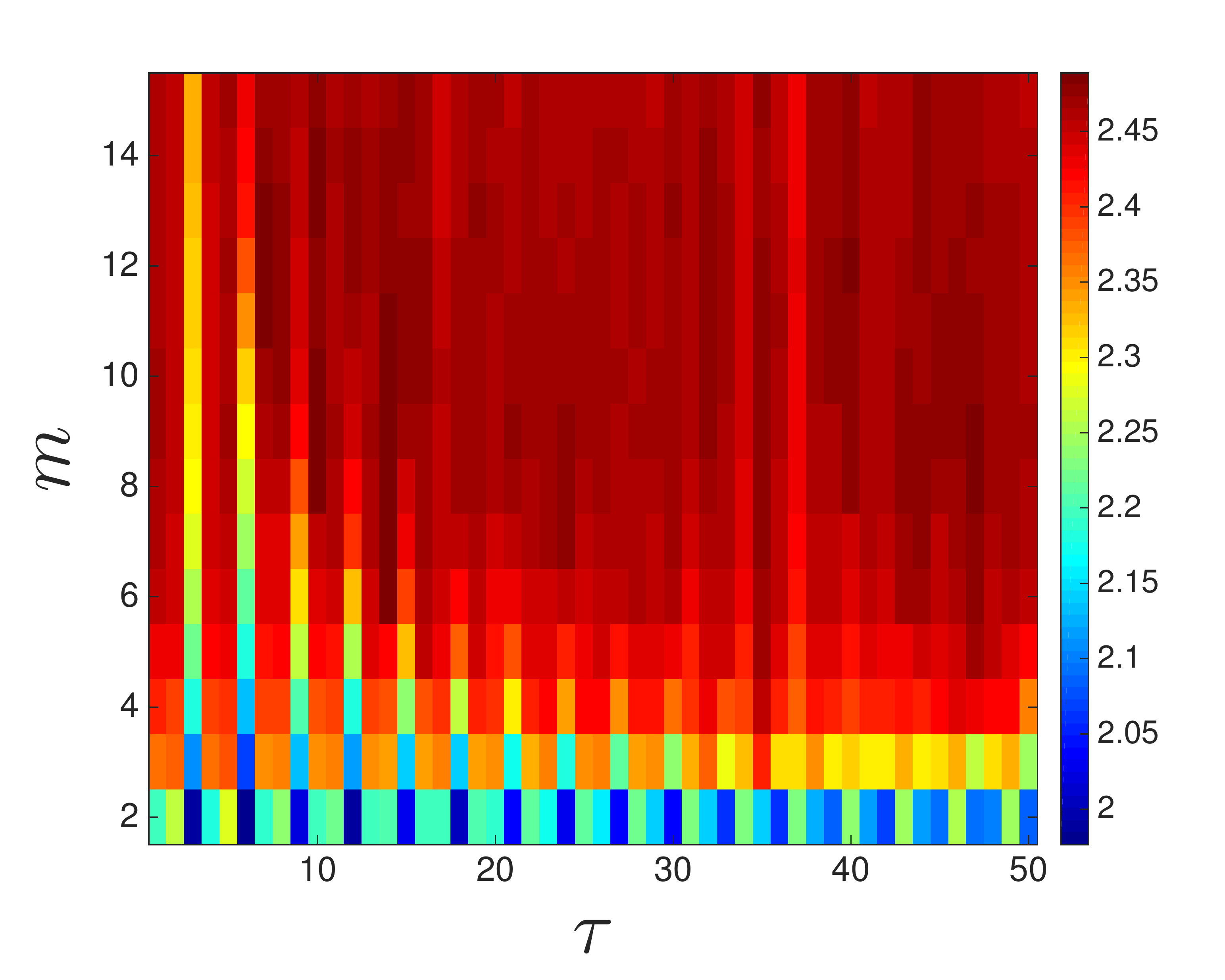}
                \caption{\mytau values for different delay
                  reconstructions of a \col trace.}
                \label{fig:colMASE}
        \end{subfigure}%
        ~ %add desired spacing between images, e. g. ~, \quad, \qquad etc.

        \begin{subfigure}[b]{\columnwidth}
                \includegraphics[width=0.9\columnwidth]{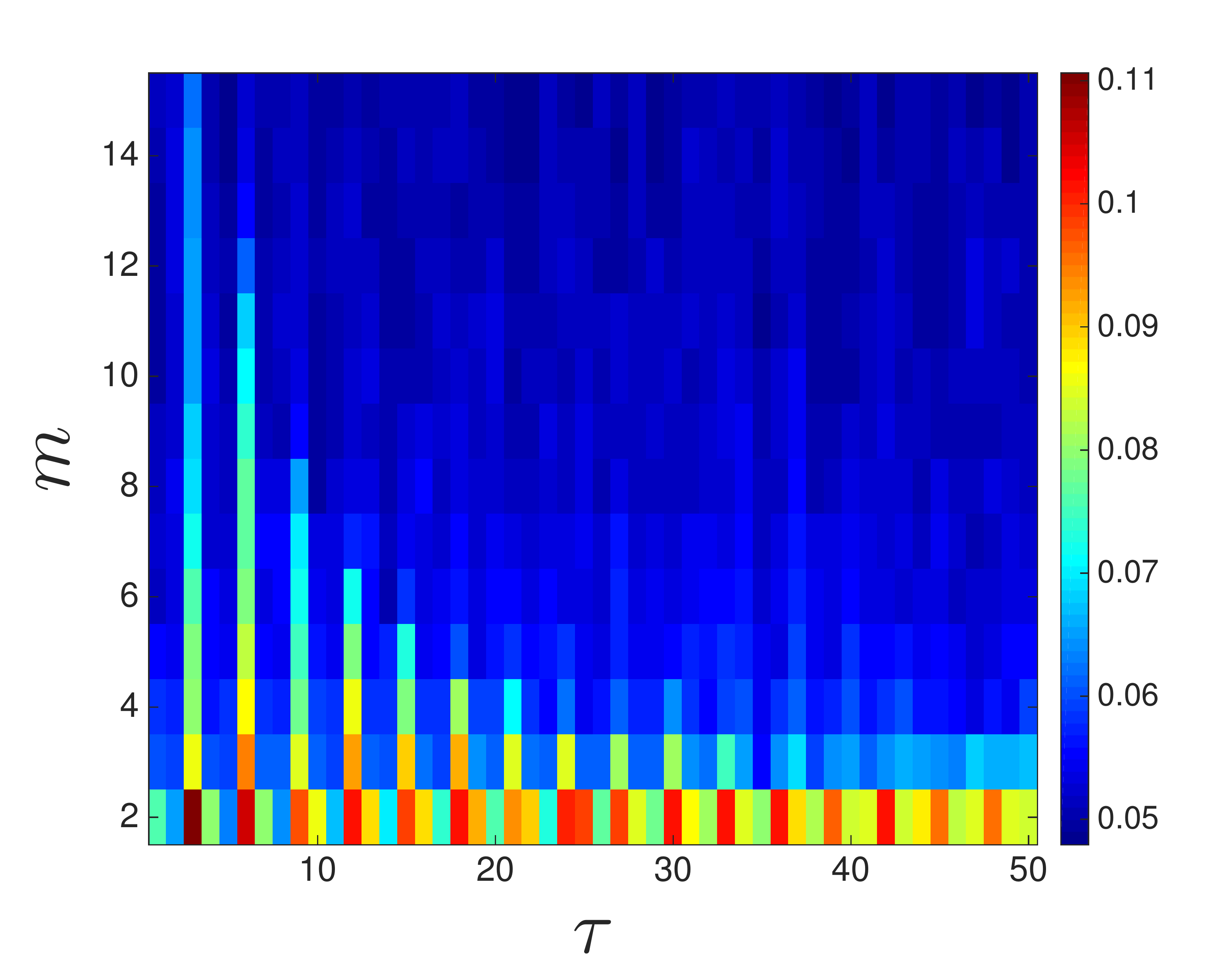}
                \caption{$MASE$ scores for LMA forecasts on different
                  delay reconstructions of a \col trace.}
                \label{fig:colmytau}
        \end{subfigure}
        %\caption{(b) Minimum MASE occurs at 0.0479 with $m=15$ and $\tau=22$ Max SPI of n.nnnn occurs at $m=?$ and $\tau=?$ .}
                \caption{The effects of reconstruction parameter
                  values on \mytau and forecast accuracy for a
                  representative trace from a computer-performance
                  dynamics experiment tracing the processor load
                  during the execution of a simple program that
                  repeatedly initializes a matrix in column-major
                  order.}
\label{fig:tauandmcol}
\end{figure}

As in the previous examples, heatmaps of $MASE$ and \mytau for the
\col time series (Figure~\ref{fig:colmytau}) are largely
antisymmetric.  
And again, reconstructions using the \mytau-optimal parameter values
allowed LMA to produce highly accurate forecasts of this signal:
$MASE_{SPI}=0.0509$, compared to the optimal $MASE_E=0.0496$.  There
are several major differences between these plots and the previous
ones in this paper, though, beginning with the vertical stripes.
These are due to the dominant unstable periodic orbit of period 3 in
the chaotic attractor in the \col dynamics.  When $\tau$ is a multiple
of this period ($\tau=3\kappa$), 
\label{page:3k} the coordinates of the delay
vector are not independent, which lowers \mytau and makes forecasting
more difficult.  (There is a nice theoretical discussion of this
effect in \cite{sauer91}.)
% 
% It is also interesting that reconstructions built with $\tau$s that
% are higher multiples of this period---e.g., 6 and 9---actually contain
% more information than the original period \alert{Not sure I understand
%   that wording.  And can you show me what you mean on the plot?  Blues
%   are getting lighter and lighter.}  This, too, corroborates the
% discussion in \cite{sauer91}.  
% 
Conversely, \mytau spikes and $MASE$ plummets when $\tau=3\kappa-1$, since
the coordinates in such a delay vector cannot share any prime factors
with the period of the orbit.  The band along the bottom of both
images is, again, due to a combination of overfolding and projection.

Another difference between the \col heatmaps and the ones in
Figures~\ref{fig:tauandmL96}, \ref{fig:tauandmL63},
and~\ref{fig:MASEandmyTAULaser} is the apparent overall trend: the
``good'' regions (low $MASE$ and high \mytau) are in the lower-left
quadrants of those heatmaps, but in the upper-right quadrant of
Figure~\ref{fig:tauandmcol}.  This is partly an artifact of the
difference in the color-map scale, which was chosen here to bring out
some important details of the structure, and partly due to that
structure itself.  Specifically, the optima of the \col heatmaps---the
large dark red and blue regions in Figures~\ref{fig:colMASE}
and~\ref{fig:colmytau}, respectively---are much broader than the ones
in the earlier sections of this paper, perhaps because the signal is
so close to periodic.  (This was also the case to some extent in the
SFI A example, for the same reason.)  This geometry makes precise
comparisons of \mytau-optimal and $MASE$-optimal parameter values
somewhat problematic, as the exact optima on two almost-flat but
slightly noisy landscapes may not be in the same place.  Indeed, the
\mytau values at $\{m_{\mytau},\tau_{\mytau}\}$ and $\{m_E,\tau_E\}$
were within a standard error across all 15 traces of \col.
% 
% \alert{$SPI_{mase_{min}}=2.4909\pm0.0100$ (std
%   error),$SPI_{MAX}=2.4961\pm0.0097$... so the max spi and the spi
%   that resulted in the min mase are within a stardard error for 15
%   samples of each other could be completely explained by statistical
%   errors in the calculations.}.
% 

\begin{figure}[ht!]
        \centering
        \begin{subfigure}[b]{\columnwidth}
                \includegraphics[width=\columnwidth]{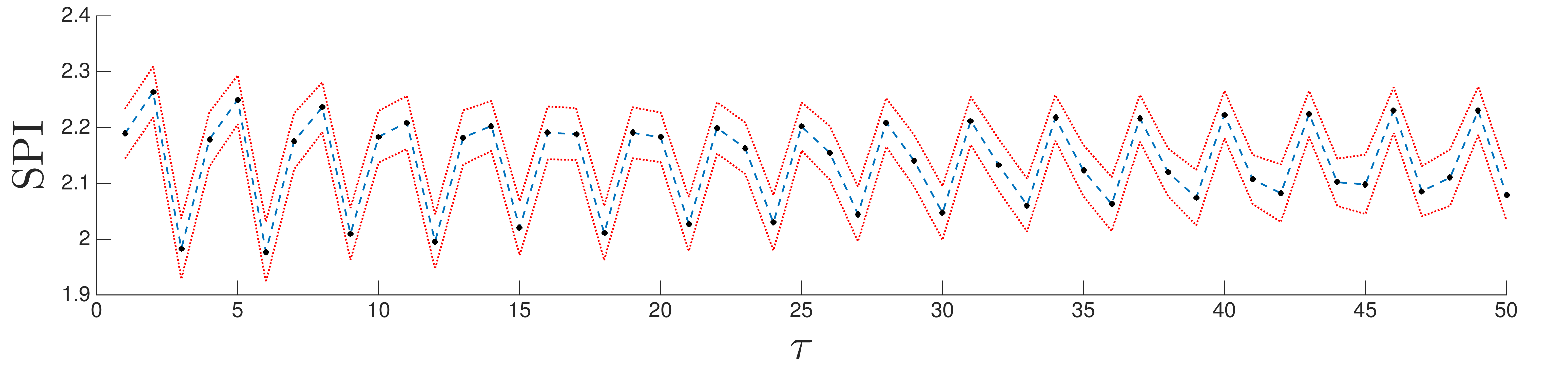}
                \caption{\mytau values for delay reconstructions of
                  the \col traces with $m=2$ and a range of values of
                  $\tau$.}
                \label{fig:colM2mytau}
        \end{subfigure}
        ~ %add desired spacing between images, e. g. ~, \quad, \qquad etc.
%          %(or a blank line to force the subfigure onto a new line)

        \begin{subfigure}[b]{\columnwidth}
                \includegraphics[width=\columnwidth]{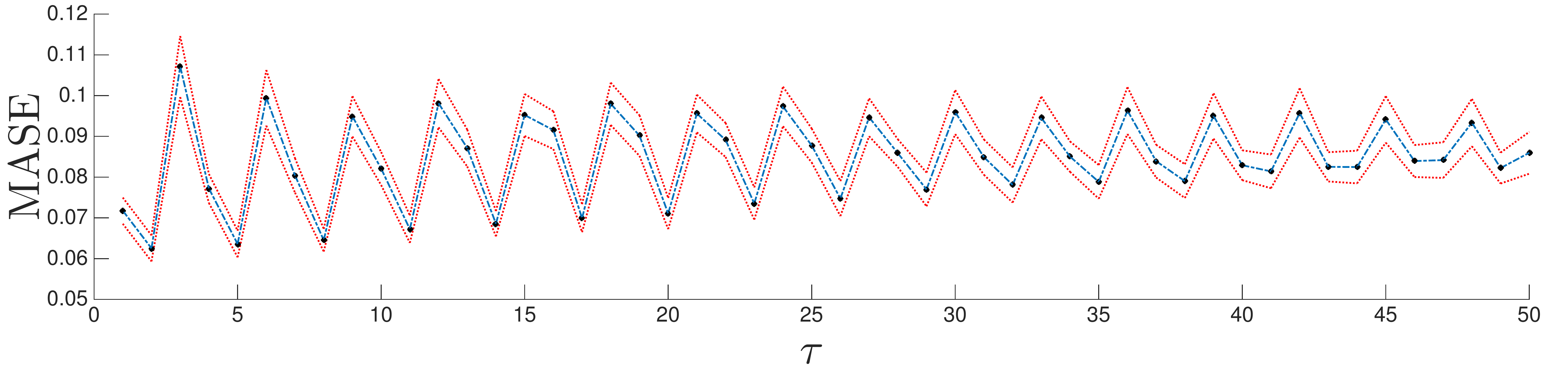}
                \caption{$MASE$ scores for LMA forecasts of delay
                  reconstructions of the \col traces with $m=2$ and a
                  range of values of $\tau$. }
                \label{fig:L96N47F5MASE}
        \end{subfigure}%
           \caption{$MASE$ and \mytau for LMA forecasts of $m=2$ delay
             reconstructions of all 15 \col traces, plotted as a
             function of $\tau$.  The blue dashed curves show the
             averages across all trials; the red dotted lines are that
             average $\pm$ the standard deviation.}
% The black \alert{maybe an arrow on the horizontal axis would work
%   better?} marks the $\tau$ that is the first minimum of the
% time-delayed mutual information curve for that time
% series. \alert{Note: that line, or arrow, needs to be on both plots}
\label{fig:MASEandmyTAUcol}
\end{figure}

And that brings up an interesting tradeoff.  For practical purposes,
what one wants is $\{m_{\mytau},\tau_{\mytau}\}$ values that produce a
$MASE$ value that is \emph{close to} the optimum $MASE_E$.  However,
the algorithmic complexity of most nonlinear time-series analysis and
prediction methods scales badly with $m$.  In cases where the \mytau
maximum is broad, then, one might want to choose the lowest value of
$m$ on that plateau---or even a value that is on the \emph{shoulder}
of that plateau, if one needs to balance efficiency over accuracy.
Indeed, forecasts with $m=2$ appear to work surprisingly well for many
nonlinear dynamical systems, including the \col data\cite{joshua-pnp}.
Fixing $m=2$ amounts to marginalizing the heatmaps in
Figure~\ref{fig:tauandmcol}, which produces a cross section like the
ones shown in Figure~\ref{fig:MASEandmyTAUcol}.
The antisymmetry between \mytau and $MASE$ is quite apparent in these
plots; the global maximum of the former coincides with the global
minimum of the latter, at $\tau=2$.  The average $MASE$ score of \col
forecasts constructed with $m=2$ and this $\tau$ value is $0.0649$.
% \pm0.003$
This is not much lower than the overall optimum of 0.0496---a value
from a forecast whose free parameters required almost six orders of
magnitude more CPU time to compute.  As an important aside: these
results suggest that one could bypass even more of the computational
effort that is involved in delay reconstruction-based forecasting by
simply working in two dimensions, i.e., by calculating \mytau across a
range of $\tau$s, rather than across a 2D $\{m,\tau\}$ space.  This
approach is discussed further in \cite{joshua-pnp}.

The correspondence between $MASE$ and \mytau also holds true for other
marginalizations: i.e., the minimum $MASE$ and the maximum \mytau
occur at the same $\tau$ value for all $m$-wise slices of the \col
heatmaps, to within statistical fluctuations.  The methods of
\cite{fraser-swinney} and \cite{KBA92}, incidentally, suggest
$\tau_H=2$ and $m_H=12$ for these traces; the $MASE$ of an LMA
forecast on such a reconstruction is 0.0530, which is somewhat better
than the best result from the $m=2$ marginalization, although still
short of the overall optimum.  The correspondence between $\tau_H$ and
$\tau_{\mytau}$ is coincidence; for this particular signal, maximizing
the independence of the coordinates happened to maximize the
information about the future contained in each delay vector.  The
$m=12$ result is not coincidence---and quite interesting, in view of
the fact that the $m=2$ forecast is so good.  It is also surprising in
view of the huge number of transistors---potential state
variables---in a modern computer.  As described in \cite{mytkowicz09},
however, the hardware and software constraints in these systems
confine the dynamics to a much lower-dimensional manifold.  All of
these issues, and their relation to the task of prediction, are
explored in more depth in \cite{joshua-pnp}.

% \gcc provides a very interesting example. It has been shown in the
% literature\cite{josh-pre} that this time series has little to no
% information sharing with the future, and this is corroborated with
% \mytau. Averaged over all choices of $m$ and $\tau$, \mytau was
% $0.5730\pm.0612$. Compare this to any of the other signals we have
% examined so far where average \mytau is much higher, e.g., 4.5 in the
% Lorenz 96 $K=22$ example above. What we would expect from this low
% \mytau and from the previous literature is that regardless of
% parameter choice LMA would perform poorly on this signal. This is
% indeed what we see, regardless of parameter choices LMA is unable to
% forecast lower than a MASE of 1, in fact average MASE was
% $1.4098\pm0.0583$. A MASE$>1$ suggests that a random walk forecast
% would have out performed the given method, i.e., all LMA methods did
% worse---significantly worse in some cases---than simply predicting the
% prior value for the next one. What this suggests is that when \mytau
% is very low for all parameter values another class of forecasting
% algorithms may be more successful and it may be the case that LMA is
% not appropriate for any choice of $m$ and $\tau$. See \cite{josh-pre}
% for a full discussion of this.

The \col program is what is known in the computer-performance
literature as a ``micro-kernel''---a extremely simple example that is
used in proof-of-concept testing.  The fact that its dynamics are so
rich speaks to the complexity of the hardware (and the
hardware-software interactions) in modern computers; again, see
\cite{todd-phd,mytkowicz09} for a much deeper discussion of these
issues.  Modern computer programs are far more complex than this
simple micro-kernel, of course, which begs the question: what does
\mytau tell us about the dynamics of truly complex systems like
that---programs that the computer-performance community models as
stochastic systems?

\begin{figure}[ht!]
        \centering
        \begin{subfigure}[b]{\columnwidth}
                \includegraphics[width=0.9\columnwidth]{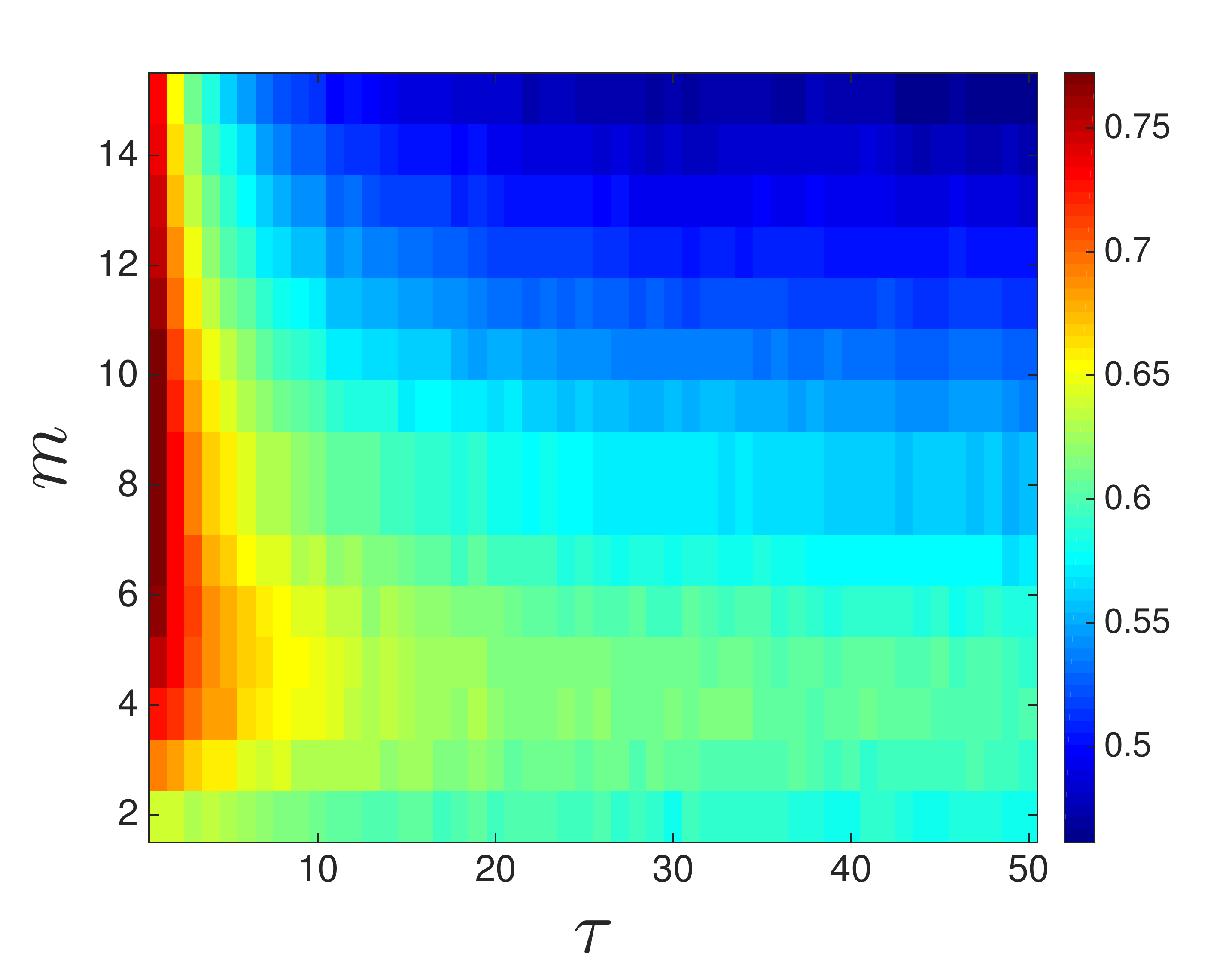}
                \caption{\mytau values for different delay
                  reconstructions of a \gcc trace.}
                \label{fig:gccmytau}
        \end{subfigure}%
        ~ %add desired spacing between images, e. g. ~, \quad, \qquad etc.

        \begin{subfigure}[b]{\columnwidth}
                \includegraphics[width=0.9\columnwidth]{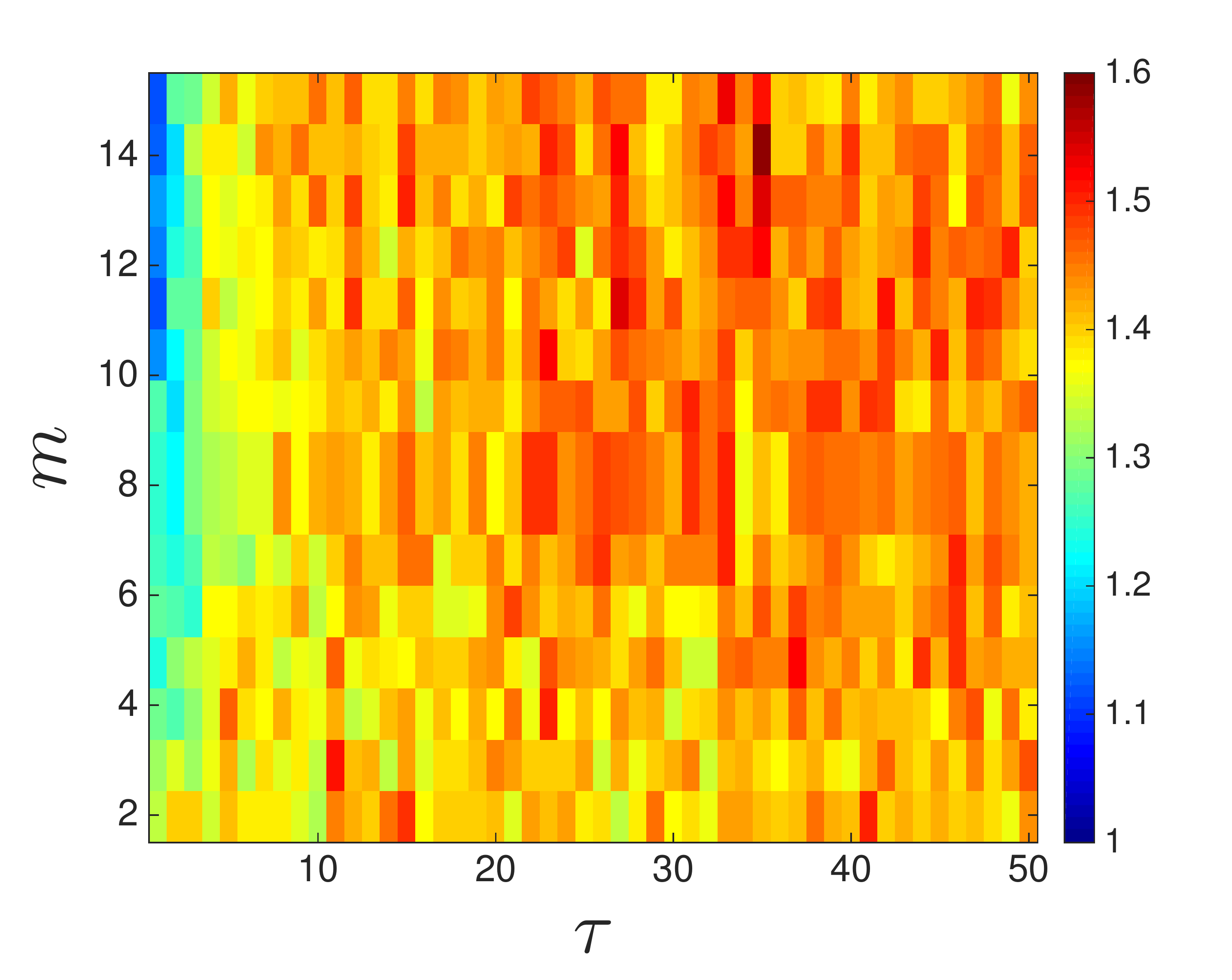}
                \caption{$MASE$ scores for LMA forecasts on different
                  delay reconstructions of a \gcc trace.}
                \label{fig:gccMASE}
        \end{subfigure}
                \caption{The effects of reconstruction parameter
                  values on \mytau and forecast accuracy for a
                  representative trace from a computer-performance
                  dynamics experiment using the \gcc benchmark}
\label{fig:tauandmgcc}
\end{figure}

For \gcc, the answer is, again, that \mytau appears to be an effective
and efficient way to assess predictability.  It has been
shown\cite{josh-pre} that this time series shares little to no
information with the future: i.e., that it \emph{cannot} be predicted
using delay reconstruction-based forecasting methods, regardless of
$\tau$ and $m$ values.  The experiments in \cite{josh-pre} required
dozens of hours of CPU time to establish that conclusion; \mytau gives
the same results in a few seconds, using much less data.  The
structure of the heatmaps for this experiment, which are shown in
Figure~\ref{fig:tauandmgcc}, is radically different.
The patterns visible in the previous $MASE$ plots, and the
antisymmetry between \mytau and $MASE$ plots, are absent from
Figure~\ref{fig:tauandmgcc}, reflecting the lack of predictive content
in this signal.  Note, too, that the color maps are different in this
Figure.  This reflects the much lower values of \mytau for this
signal: a maximum \mytau of 0.7722 for \gcc, compared to 5.3026 for
Lorenz 96 with $K=22$.  
% 
% \alert{Add some sort of MASE range comparison?}  
% 
Indeed, the $MASE$ surface in Figure~\ref{fig:gccMASE} never dips
below 1.0\footnote{Figure~\ref{fig:tauandmL96}, in contrast, never
  exceeds $\approx 0.6$ and generally stays below 0.3.}.  That is,
regardless of parameter choice, LMA forecasts of \gcc are no better
than simply using the prior value of this scalar time series as the
prediction.  The uniformly low \mytau values in
Figure~\ref{fig:gccmytau} are an effective indicator of this---and,
again, they can be calculated quickly, from a relatively small sample
of the data.  It is to that issue that we turn next.

\section{Data Requirements and Prediction Horizons}
\label{sec:dataandhorizon}

In some real-world situations, it may be impractical to rebuild
forecast models at every step, as we have done in the previous
sections of this paper---because of computational expense, for
instance, or because the data rate is very high.  In these situations,
one may wish to predict $p$ time steps into the future, then stop and
rebuild the model to incorporate the $p$ points that have arrived
during that period, and repeat.  In chaotic systems, of course, there
are fundamental limits on prediction horizon even if one is working
with infinitely long traces of all state variables.  A key question at
issue in this section is how that effect plays out in forecast models
that use delay reconstructions from scalar time-series data.  And
since real-world data sets are not infinitely long, it is important to
understand the effects of data length on the estimation of \mytau.

\subsection{Data Requirements for \mytau Estimation}
\label{subsec:datalength}

% One nice feature of using \mytau for parameter selection is that it
% allows a practitioner to decide how many reconstruction dimensions are
% appropriate directly from the data that is available. 

The quantity of data used in a delay reconstruction directly impacts
the usefulness of that reconstruction.  If one is interested in
approximating the correlation dimension via the Grassberger-Procaccia
algorithm, for instance, it has been shown that one needs
$10^{(2+0.4m)}$ data points\cite{tsonisdatabound,smithdatabound}.
Those bounds are overly pessimistic for forecasting, however.
% For example, Sauer~\cite{sauer-delay} successfully forecasted the
% continuation of a 16,000-point time series (interpolated version of
% Data Set A from Section~\ref{sec:intro}) embedded in 16 dimensions;
% Sugihara \& May~\cite{sugihara90} used delay-coordinate embedding with
% $m$ as large as seven to successfully forecast biological and
% epidemiological time-series data as short as 266 points, both of which
% would have needed orders of magnitude more data according to
% \cite{tsonisdatabound,smithdatabound}.
For example, Sugihara \& May~\cite{sugihara90} used delay-coordinate
reconstructions with $m$ as large as seven to successfully forecast
biological and epidemiological time-series data sets that contain as
few as 266 points.  A key challenge, then, is to determine whether
one's time series \emph{really} calls for as many dimensions and data
points as the theoretical results require, or whether one can get away
with fewer dimensions---and how much data one needs in order to figure
all of that out.

We claim that \mytau is a useful solution to those challenges.  As
established in the previous sections, calculations of this quantity
can reveal what dimension one needs to build a good delay
reconstruction for the purposes of LMA forecasting of nonlinear and
chaotic systems.  And, as alluded to in those sections, \mytau can be
estimated accurately from a surprisingly small number of points.  The
experiments in this section explore that intertwined pair of claims in
more depth by increasing the length of the Lorenz 96 traces and
testing whether the information content of the state estimator derived
from standard heuristics converges to the \mytau-optimal
estimator\footnote{This kind of experiment is not possible in
  practice, of course, when the time series is fixed, but can be done
  in the context of this synthetic example.}.
%It should be noted that this data-length effect is also paralleled
%when estimating a higher-dimensional probability distribution
%function. In the case of \roLMA I need to estimate a 3-dimensional
%distribution, whereas with \fnnLMA I need to estimate a 9-dimensional
%distribution. The latter requiring significantly more
%data. \alert{basically here I am trying to say that yes I see that
%\mytau could be less because of this high dimensional pdf estimation
%but this is a real effect in both \fnnLMA as well as \mytau and so
%while it is a concern it is a parallel concern...I have no idea how
%to say this...}

\begin{figure}[b!]
        \centering
                \begin{subfigure}[b]{\columnwidth}
                \includegraphics[width=\columnwidth]{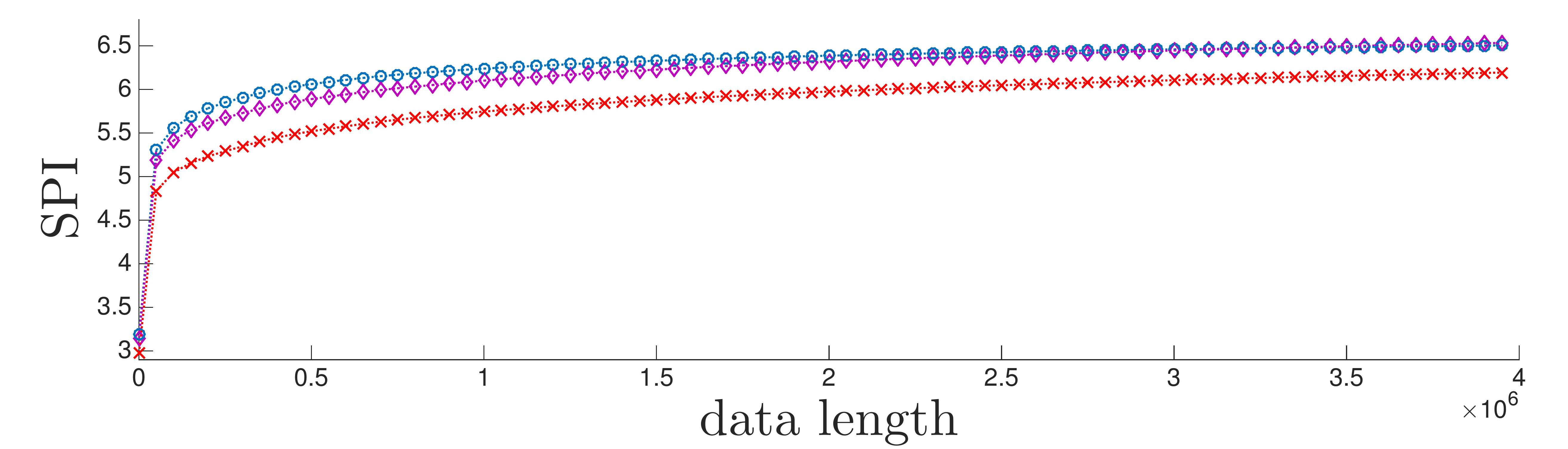}
                \caption{Optimal \mytau for traces from the
                  $\{K=22,F=5\}$ Lorenz 96 system}
                \label{fig:data22}
        \end{subfigure}

        \begin{subfigure}[b]{\columnwidth}
                \includegraphics[width=\columnwidth]{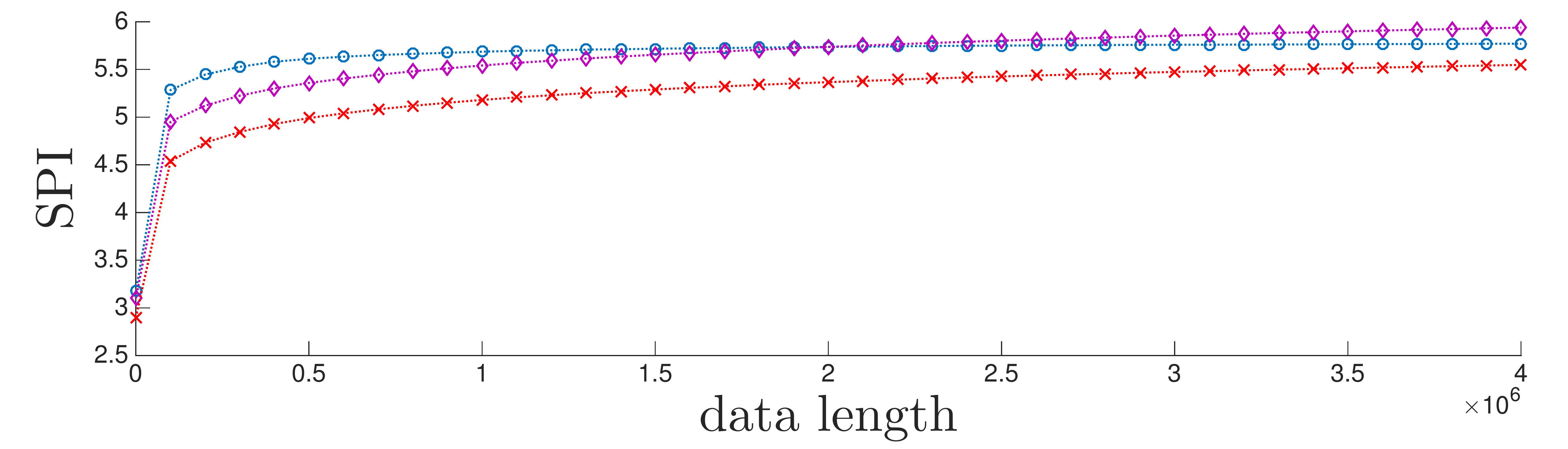}
                \caption{Optimal \mytau for traces from the
                  $\{K=47,F=5\}$ Lorenz 96 system}
               % \label{fig:gccMASE}
        \end{subfigure}%
                  \caption{\mytau versus data length for
                    traces from the Lorenz-96 system using $\tau=1$ in all cases. Blue circles
                   corresponds to an embedding dimension $m=2$,
                    purple diamonds to $m=4$, and red xs to $m=8$. }\label{fig:mytaudata}
\end{figure}

Figure~\ref{fig:mytaudata} shows the results.
% we have increased Lorenz 96 (a) $K=22, F=5$
% and (b) $K=47,F=5$, time series from 500 to 4 million points and
% calculated \mytau for the optimal $\tau$ (as determined by \mytau)
% value ($\tau=1$) for $m=2, 4$ and 8 over 15 different trajectories.
When the data length is short, the $m=2$ state estimator had the most
information about the future.  This makes perfect sense; a short time
series cannot fully sample a complicated object, and when an
ill-sampled high-dimensional manifold is projected into a low
dimensional space, infrequently visited regions of that manifold can
act effectively like noise.  From an information-theoretic standpoint,
this would increase the effective Shannon entropy rate of each of the
variables in the delay vector.  In the I-diagram in
Figure~\ref{fig:mytau}, this would manifest as drifting apart of the
two circles, decreasing the shaded region that one needs to maximize
for effective forecasting.

If that reasoning is correct, longer data lengths should fill out the
attractor, thereby mitigating the spurious increase in the Shannon
entropy rate and allowing higher-dimensional reconstructions to
outperform lower-dimensional ones.  This is indeed what one sees in
Figure~\ref{fig:mytaudata}.  For both the $K=22$ and $K=47$ traces,
once the signal is 2 million points long, the four-dimensional
estimator has caught up to and even exceeded the two-dimensional case.
% \alert{Josh,
  %since the $K=47$ attractor is higher dimensional, shouldn't the
  %catchup happen much later?  That would be what one would expect,
  %given the argument above.}  
  Note, though, that the optimal \mytau of
the $m=8$ reconstruction model is still lower than in the $m=2$ or $4$
cases, even at the right-hand limit of the plots in
Figure~\ref{fig:mytaudata}.  That is, even with a time series that
contains $4 \times 10^6$ points, it is more effective to use a lower
dimensional reconstruction to make an LMA forecast.  But the really
important message here is that \mytau allows one to determine the best
reconstruction parameters \emph{for the available data}, which is an
important part of the answer to the challenges outlined at the
beginning of this section.

Something very interesting happens in the $m=2$ results for Lorenz 96
model with $K=47$: the \mytau curve reaches a maximum value around
100,000 points and stops increasing, regardless of data length.  What
this means is that this two-dimensional reconstruction contains as
much information about the future as can be ascertained from these
data, suggesting that increasing the length of the training set would
not improve forecast accuracy.  To explore this, we constructed LMA
forecasts of different-length traces (100,000--2.2 million points)
from this system, then reconstructed their dynamics with different $m$
values and the appropriate $\tau_{\mytau}$ for each case.  For $m=2$,
both \mytau and $MASE$ results did indeed plateau at 100,000
points---at 5.736 and 0.0809, respectively.  As before, more data does
afford higher-dimensional reconstructions more traction on the
prediction problem: the $m=4$ forecast accuracy surpassed $m=2$ at
around 2 million points ($MASE=0.0521$).  In neither case, by the way,
did $m=8$ catch up to either $m=2$ or $m=4$, even at 4 million data
points.  Of course, one must consider the cost of storing the
additional variables in a higher-dimensional reconstruction model,
particularly in data sets this long, so it may be worthwhile in
practice to settle for the $m=2$ forecast---which is only slightly
less accurate and requires only 100,000 points.  This has another
major advantage as well.  If the time series is non-stationary, a
forecast strategy that requires fewer points can adapt more quickly.

\subsection{Choosing reconstruction parameters for increased prediction horizons.}
\label{subsec:predictionhorizon}

So far in this paper, we have considered forecasts that were
constructed one step at a time and studied the correspondence of their
accuracy with one-step-ahead calculations of \mytau.  In this section,
we consider longer prediction horizons ($p$) and explore whether one
can use a $p$-step-ahead version of \mytau---i.e.,
$I[\mathcal{S}_j,X_{j+p}]$, with $p>1$---to choose parameter values
that maximize the information contained in each delay vector about the
value of the time series $p$ steps in the future.
% Own up to fact that We do not make direct comparison with MASE
% & explain why - mase is over the whole range whereas SPI only looks
% p steps out (and not in between)

\begin{figure}[ht!]
        \centering
                \includegraphics[width=\columnwidth]{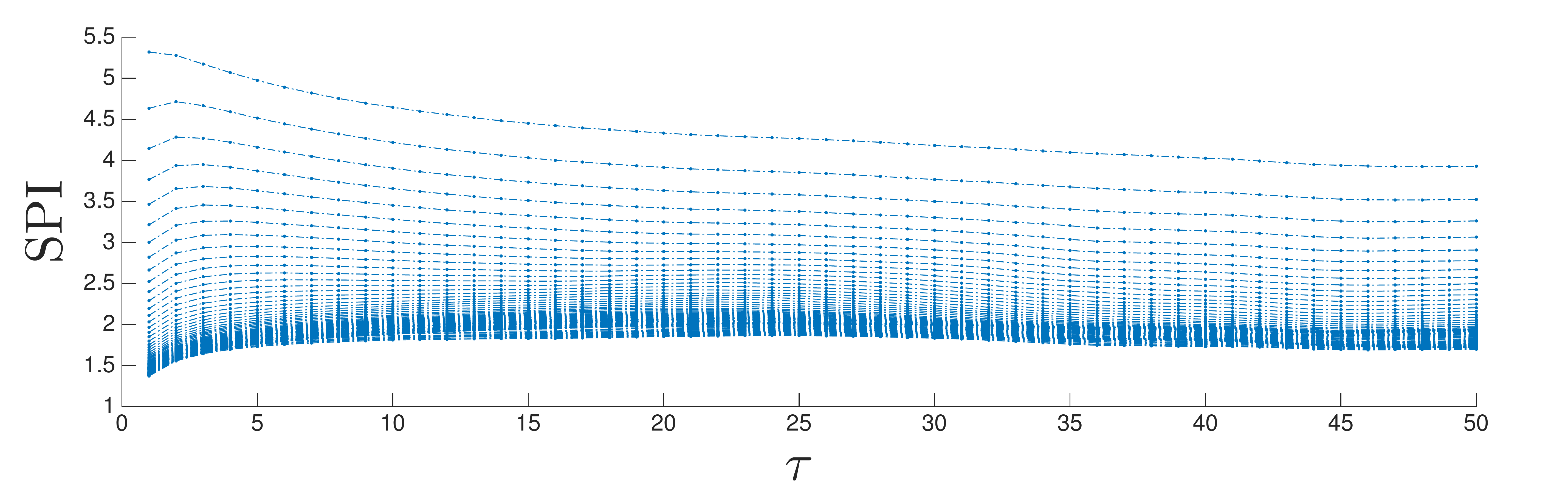}
           \caption{The effect of prediction horizon ($p$) on \mytau
             of the $K=22$ Lorenz 96 system for a fixed reconstruction
             dimension ($m=2$).  The traces in the image, starting
             from the top, correspond to prediction horizons of $p=1$
             to $p=100$.}
           \label{fig:L22pvsTau}
\end{figure}

One would expect the \mytau-optimal $\{m,\tau\}$ values for a given
time series to depend on the prediction horizon.  It has been shown,
for instance, that longer-term forecasts generally do better with
larger $\tau$\cite{kantz97}, and conversely\cite{joshua-pnp}.  It
makes sense that one might need to reach different distances into the
past (via the span of the delay vector) in order to reduce the
uncertainty about events that are further into the
future\cite{weigend-book}.  These effects are corroborated by \mytau.
Figure~\ref{fig:L22pvsTau} demonstrates this in the context of the
Lorenz 96 system with $K=22$, focusing on $m=2$ for simplicity.
The topmost trace in this figure is for the $p=1$ case---i.e., a
horizontal slice of Figure~\ref{fig:L96N22F5SPI} made at $m=2$.  The
maximum of this curve is the optimal $\tau$ value ($\tau_{\mytau}$)
for this reconstruction.  The overall shape of this trace reflects the
monotonic increase in the uncertainty about the future with $\tau$
that is noted on page~\pageref{page:increase-with-tau}.  The other
traces in Figure~\ref{fig:L22pvsTau} show \mytau as a function of
$\tau$ for $p=2, 3, \dots$, down to $p=100$ at the bottom of the
figure.  The lower traces do not decrease monotonically; rather, there
is a slight initial rise.  This is due to the point made above about
the span of the delay vector: if one is predicting further into the
future, it may be useful to reach further into the past.  In general,
this causes the optimal $\tau$ to shift to the right as prediction
horizon increases, going down the plot---i.e., longer prediction
horizons require a greater $\tau$ (cf. \cite{kantz97}).  For very long
horizons, the choice of $\tau$ appears to matter very little.  In
particular, \mytau is fairly constant and quite low for $5<\tau<50$
when $p>30$---i.e., regardless of the choice of $\tau$, there is very
little information about the $p$-distant future in any delay
reconstruction of this signal for $p>30$.  This effect should not be
surprising, and it is well corroborated in the literature.  However,
it can be hard to know {\sl a priori}, when one is confronted with a
data set from an unknown system, to know what prediction horizon makes
sense.  \mytau offers a computationally efficient way to answer that
question.

\begin{figure}[b!]
        \centering
        %\begin{subfigure}[b]{\textwidth}
                \includegraphics[width=\columnwidth]{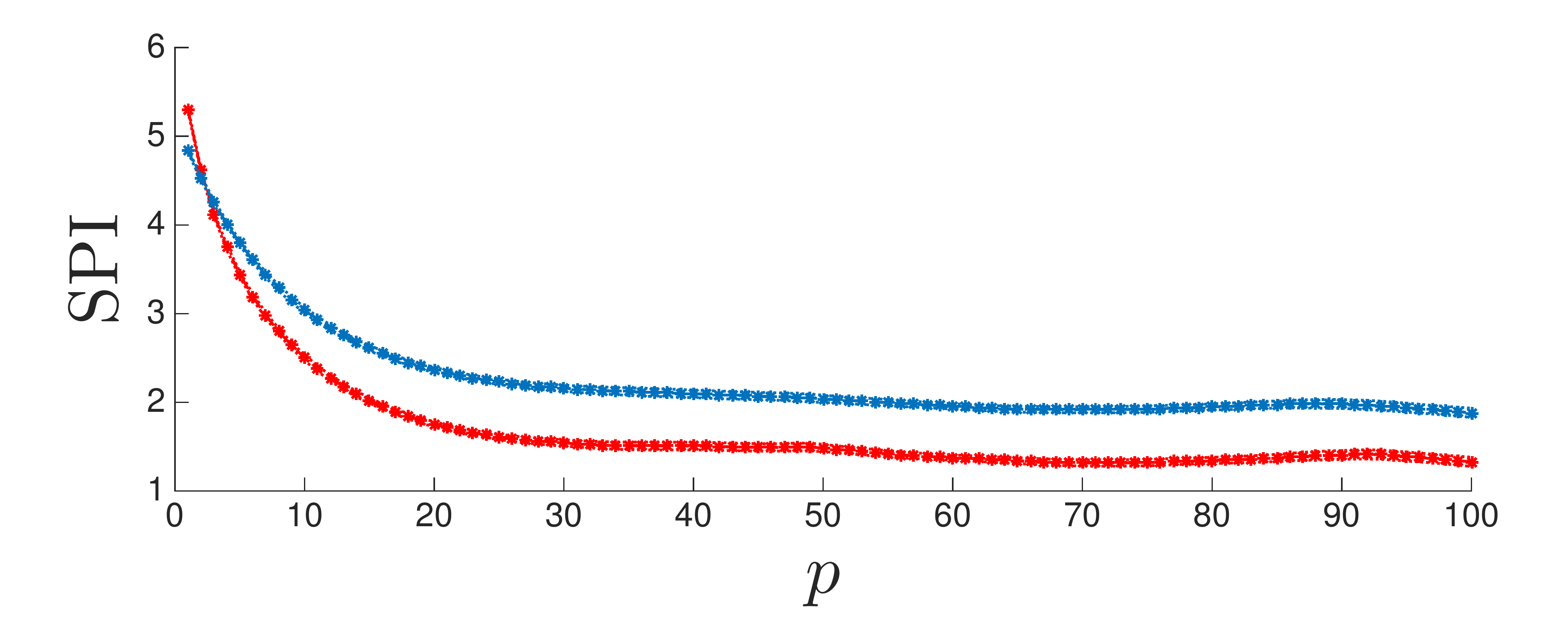}
                %\caption{red = \roLMA, blue = \fnnLMA}
                %\label{fig:L22mytauhorizon}
        %\end{subfigure}%
         %add desired spacing between images, e. g. ~, \quad, \qquad etc.

        %\begin{subfigure}[b]{\textwidth}
        %        \includegraphics[width=\textwidth]{figs/gccpHorizon}
        %        \caption{red = \roLMA, blue = \fnnLMA }
        %        \label{fig:gccmytauhorizon}
        %\end{subfigure}
            \caption{The effect of prediction horizon ($p$) on \mytau
              of the $K=22$ Lorenz 96 system for a fixed time delay
              ($\tau=1$) and two different reconstruction dimensions.
              The red line is $m=2$ and the blue is
              $m_H=8$, the value suggested for this signal
              by the technique of false neighbors.}
           \label{fig:L22mytauhorizon}
\end{figure}

Figure~\ref{fig:L22mytauhorizon} shows a similar exploration of the
other side of that question: the effects of the reconstruction
dimension on \mytau, with $\tau$ fixed at 1.
The $m=2$ state estimator contains more information about the future
for short prediction horizons.  This ties back to the discussion at
the end of Section~\ref{subsubsec:computer}: low-dimensional
reconstructions can work quite well for short prediction horizons.
However, Figure~\ref{fig:L22mytauhorizon} shows that the full
reconstruction is better for longer horizons.  This is not terribly
surprising, since a higher reconstruction dimension allows the state
estimator to capture more information about the past.  Finally, note
that \mytau decreases monotonically with prediction horizon for both
$m=2$ and $m_H$.  This, too, is unsurprising.  Pesin's
relation\cite{pesin1977characteristic} says that the sum of the
positive Lyapunov exponents is equal to the entropy rate, and if there
is a non-zero entropy rate, then generically observations will become
increasingly independent the further apart they are.  This explanation
also applies to Figure~\ref{fig:L22pvsTau}, of course, but it does
\emph{not} hold for signals that are wholly (or nearly) periodic.

%Similarly in Table~\ref{} we see from the $p$-MASE scores for \roLMA and \fnnLMA for forecasting \gcc gradually decrease together and by $h=50$ the forecasts are quite bad, only doing one order of magnitude better than a 50 step ahead random walk forecaster!  This is reflected by the \mytau as well. By $h=50$ the \mytau has dropped to near 0.2, which is effectively no reduction in uncertainty about the future, this means that both LMA based approaches are basically guessing what to forecast and getting lucky more than random walk.

%\begin{figure}[ht!]
%        \centering
        %\begin{subfigure}[b]{\textwidth}
        %        \includegraphics[width=\textwidth]{figs/Lorenz22pHorizon}
        %        \caption{red = \roLMA, blue = \fnnLMA}
        %        \label{fig:L22mytauhorizon}
        %\end{subfigure}%
        %~ %add desired spacing between images, e. g. ~, \quad, \qquad etc.
        %  %(or a blank line to force the subfigure onto a new line)
        %
        %\begin{subfigure}[b]{\textwidth}
%                \includegraphics[width=\columnwidth]{figs/gccpHorizon}
                %\caption{red = \roLMA, blue = \fnnLMA }
                %\label{fig:gccmytauhorizon}
        %\end{subfigure}
%           \caption{\gcc, red = \roLMA, blue = \fnnLMA }\label{fig:gccmytauhorizon}
%\end{figure}

\begin{figure}[b!]
        \centering
                \includegraphics[width=\columnwidth]{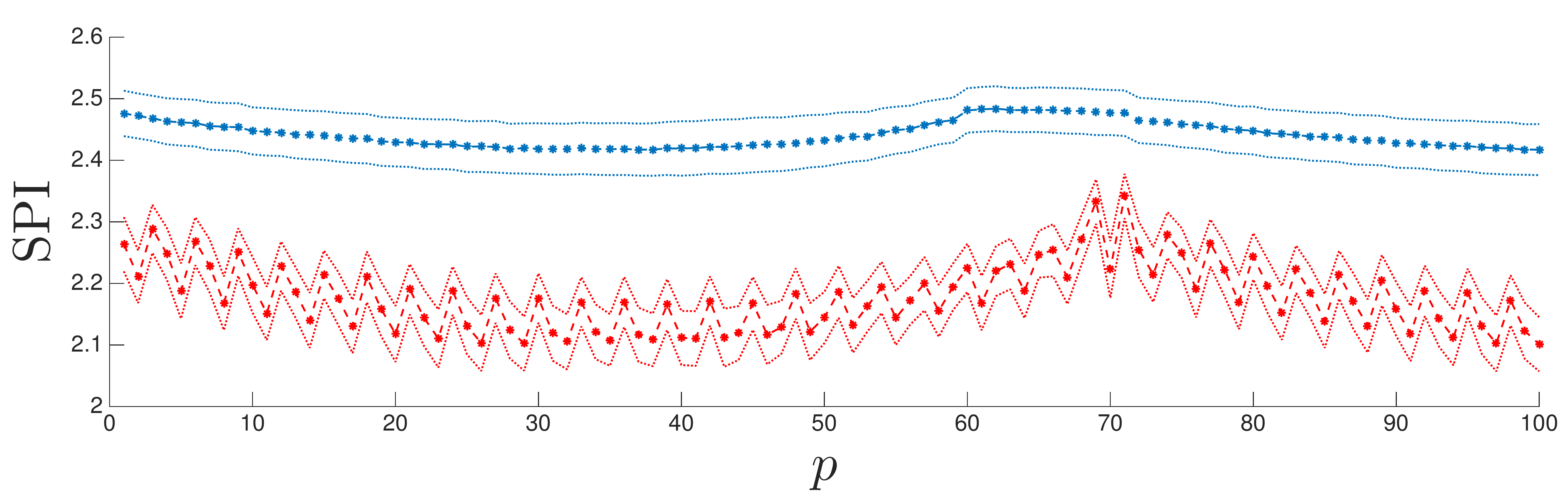}
            \caption{The effect of prediction horizon ($p$) on \mytau
              of the \col for a fixed time delay ($\tau=1$) and two
              different reconstruction dimensions.  The red line is
              $m=2$ and the blue is $m_H=12$, the value
              suggested for this signal by the technique of false
              neighbors.}\label{fig:colmytauhorizon}
\end{figure}

Recall that the \col dynamics in Section~\ref{subsubsec:computer} were
chaotic, but with a dominant unstable periodic orbit---which had a
variety of interesting effects in the results.
Figure~\ref{fig:colmytauhorizon} explores the effects of prediction
horizon on those results.
Not surprisingly, there is some periodicity in the \mytau versus $p$
relationships, but not for the same reasons that caused the stripes in
Figure~\ref{fig:colmytau}.  Here, the \emph{peaks} in \mytau occur at
multiples of the period.  That is, the $m=2$ state estimator can
forecast with the most success when the value being predicted is in
phase with the delay vector.  Note that this effect is far stronger
for $m=2$ than $m_H$, simply because of the instability of that
periodic orbit; the visits made by the chaotic trajectory to that
orbit are more likely to be short than long.  As expected, \mytau
decays with prediction horizon---but only at first, after which it
begins to rise again, peaking at $p=69$ and $p=71$.  This may be due
to a second higher-order unstable periodic orbit in the \col dynamics.

In theory, one can derive rigorous bounds on prediction horizon.  The
time at which $\mathcal{S}_j$ will no longer have any information
about the future can be determined by considering:
\begin{align*}
  R(p) = \frac{I[\mathcal{S}_j;X_{j+p}]}{H[X_{j+p}]}~,
\end{align*}
i.e., the percentage of the uncertainty in $X_{j+p}$ that can be
reduced by the delay vector.  Generically, this will limit to some
small value equal to the amount of information that the delay vector
contains about any arbitrary point on the attractor.  Given some
criteria regarding how much information above the ``background'' is
required of the state estimator, one could use the $R(p)$ versus $p$
to determine the maximum practical horizon.

In practice, one can select parameters for delay reconstruction-based
forecasting by explicitly including the prediction horizon in the
\mytau function, fixing its value at the required value, performing
the same search as we did in earlier sections over a range of $m$ and
$\tau$, and then choosing a point on (or near) the optimum of that
\mytau surface.  The computational and data requirements of this
calculation, as shown in Section~\ref{subsec:datalength}, are far
superior to those of the standard heuristics used in delay
reconstructions.

\section{Conclusion}
\label{sec:conclusion}

In this paper, we have described a new metric for quantifying how much
information about the future is contained in a delay reconstruction.
Using a number of different dynamical systems, we demonstrated a
direct correspondence between the \mytau value for different delay
reconstructions and the accuracy of forecasts made with Lorenz's
method of analogues on those reconstructions.  Since \mytau can be
calculated quickly and reliably from a relatively small amount of
data, without needing to know anything about the governing equations
or the state space dynamics of the system, that correspondence is a
major advantage, in that it allows one to choose parameter values for
delay reconstruction-based forecast models without doing an exhaustive
search on the parameter space.  Significantly, \mytau-optimal
reconstructions are better, for the purposes of forecasting, than
reconstructions constructed using standard heuristics like mutual
information and the method of false neighbors, which can require large
amounts of data, significant computational effort, and expert human
interpretation.  \mytau allows us to answer other questions regarding
forecasting with theoreticaly unsound models\cite{joshua-pnp}---e.g.,
why one can obtain a better forecast using a low-dimensional
reconstruction than with a full embedding.  It also allows one to
understand bounds on prediction horizon without having to estimate
Lyapunov spectra or Shannon entropy rates, which are difficult to
obtain for arbitrary real-valued time series.  That, in turn, allows
one to tailor one's reconstruction parameters to the amount of
available data and the desired prediction horizon---and to know if a
given prediction task is just not possible.

The explorations in this paper involve a simple near-neighbor forecast
strategy and state estimators that are basic delay reconstructions of
raw time-series data.  The definition and calculation of \mytau do not
involve any assumptions about the state estimator, though, so the
results presented here should also hold for other state estimators.
For example, it is common in time-series prediction to pre-process
one's data: for example, low-pass filtering or interpolating to
produce additional points.  Calculating \mytau after performing such
an operation will accurately reflect the amount of information in that
new time series---indeed, it would reveal if that pre-processing step
\emph{destroyed} information.  And we believe that the basic
conclusions in this paper extend to other state-space based forecast
schemas besides Lorenz's method of analogues, such as those used in
\cite{weigend-book,casdagli-eubank92,Smith199250,sugihara90,sauer-delay}---although
\mytau may not accurately select optimal parameter values for
strategies that involve post-processing the data (e.g.,
GHKSS\cite{ghkss}).  We are in the process of exploring this.

There are many other interesting potential ways to leverage \mytau in
the practice of forecasting.  If the \mytau-optimal $\tau =1$, that
may be a signal that the time series is undersampling the dynamics and
that one should increase the sample rate.  One could use \mytau at a
finer grain to optimizing $\tau$ individually for each dimension, as
suggested in~\cite{pecoraUnified}.  To do this, one could define
$\mathcal{S}_j= [X_{j}, X_{j-\tau_1}, X_{j-\tau_2}, \dots,
  X_{j-\tau_{m-1}}]$ and then simply maximize \mytau using that state
estimator constrained over $\{\tau_i\}_{i=1}^{m-1}$.

  \section{References}
\bibliographystyle{elsarticle-num}
%\bibliographystyle{alpha}
%\bibliography{master-refs}

\end{document}